\documentclass[%
reprint,
superscriptaddress,
showpacs,
 bibnotes,
 amsmath,amssymb,
 aps,
prl
]{revtex4-1}

\usepackage{graphicx}
\usepackage{dcolumn}
\usepackage{bm}
\usepackage{color}

\begin{document}


\title{Doping-Dependent Raman Resonance in the Model High-Temperature Superconductor HgBa$_2$CuO$_{4+\delta}$}

\author{Yuan~Li}
\email[Electronic address: ]{yuan.li@pku.edu.cn}
\affiliation{International Center for Quantum Materials, School of Physics, Peking University, Beijing 100871, China}
\affiliation{Max Planck Institute for Solid State Research, D-70569 Stuttgart, Germany}
\author{M.~Le~Tacon}
\email[Electronic address: ]{m.letacon@fkf.mpg.de}
\affiliation{Max Planck Institute for Solid State Research, D-70569 Stuttgart, Germany}
\author{Y.~Matiks}
\affiliation{Max Planck Institute for Solid State Research, D-70569 Stuttgart, Germany}
\author{A.~V.~Boris}
\affiliation{Max Planck Institute for Solid State Research, D-70569 Stuttgart, Germany}
\author{T.~Loew}
\affiliation{Max Planck Institute for Solid State Research, D-70569 Stuttgart, Germany}
\author{C.~T.~Lin}
\affiliation{Max Planck Institute for Solid State Research, D-70569 Stuttgart, Germany}
\author{Lu~Chen}
\affiliation{International Center for Quantum Materials, School of Physics, Peking University, Beijing 100871, China}
\author{M.~K.~Chan}
\affiliation{School of Physics and Astronomy, University of Minnesota, Minneapolis, Minnesota 55455, USA}
\author{C.~Dorow}
\affiliation{School of Physics and Astronomy, University of Minnesota, Minneapolis, Minnesota 55455, USA}
\author{L.~Ji}
\affiliation{School of Physics and Astronomy, University of Minnesota, Minneapolis, Minnesota 55455, USA}
\author{N.~Bari\v{s}i\'{c}}
\affiliation{School of Physics and Astronomy, University of Minnesota, Minneapolis, Minnesota 55455, USA}
\affiliation{Service de Physique de l'Etat Condens\'e, CEA-DSM-IRAMIS, F-91198 Gif-sur-Yvette, France}
\author{X.~Zhao}
\affiliation{School of Physics and Astronomy, University of Minnesota, Minneapolis, Minnesota 55455, USA}
\affiliation{State Key Lab of Inorganic Synthesis and Preparative Chemistry, College of Chemistry, Jilin University, Changchun 130012, P.R. China}
\author{M.~Greven}
\affiliation{School of Physics and Astronomy, University of Minnesota, Minneapolis, Minnesota 55455, USA}
\author{B.~Keimer}
\affiliation{Max Planck Institute for Solid State Research, D-70569 Stuttgart, Germany}


\begin{abstract}
We study the model high-temperature superconductor HgBa$_2$CuO$_{4+\delta}$ with electronic Raman scattering and optical ellipsometry over a wide doping range. The resonant Raman condition which enhances the scattering cross section of ``two-magnon'' excitations is found to change strongly with doping, and it corresponds to a rearrangement of inter-band optical transitions in the 1-3 eV range seen by ellipsometry. This unexpected change of the resonance condition allows us to reconcile the apparent discrepancy between Raman and x-ray detection of magnetic fluctuations in superconducting cuprates. Intriguingly, the strongest variation occurs across the doping level where the antinodal superconducting gap reaches its maximum.

\end{abstract}

\maketitle

Magnetic fluctuations might play an essential role in the mechanism of high-temperature superconductivity in the cuprates \cite{ScalapinoPhysRep1995,*AbanovAdvPhys2003,*MoriyaRepProgPhys2003}. Upon doping the insulating parent compounds, the fluctuations evolve from antiferromagnetic (AF) magnon excitations up to a few hundred meV.  As this energy is in principle sufficient to support superconductivity at very high temperatures, the observation of similar ``paramagnon'' excitations by resonant inelastic x-ray scattering (RIXS) in superconducting cuprates \cite{LeTaconNatPhys2011} is a revealing result. Recent electronic Raman scattering (ERS) measurements further suggest that the high-energy magnetic fluctuations exhibit a pronounced change concurrent with the formation of Cooper pairs \cite{LiPRL2012}, corroborating a close connection between them.

Here we address a major puzzle that has arisen from the comparison of the doping dependent RIXS and ERS cross sections. Both techniques use inelastic photon scattering to probe fundamental excitations in solids, but with very different incident photon energies in the x-ray and visible-light range, and they can detect magnetic fluctuations in the cuprates via the creation of single- \cite{BraicovichPRL2010} and two-magnon \cite{LyonsPRB1988} excitations, respectively. In the undoped AF insulating compounds, the superexchange energies $J$ determined by RIXS and ERS agree reasonably well \cite{LyonsPRB1988,BraicovichPRL2010} and are furthermore consistent with inelastic neutron scattering (INS) measurements \cite{ColdeaPRL2001}. A comparison between these measurements at finite doping, however, reveals an important discrepancy: while the energy and spectral weight of the (para)magnon excitations observed by RIXS exhibit little change \cite{LeTaconNatPhys2011,LeTaconCondmat2013,*Deancondmat2013}, both of these quantities decrease substantially in ERS data \cite{CooperPRB1993,BlumbergPRB1994,RuebhausenPRB1997,SugaiPRB2003}. The latter observation has created the impression that the AF spin fluctuations become overdamped near and above optimal doping \cite{BlumbergPRB1994}. This, in turn, has served as a major argument against magnetically driven Cooper pairing in the overdoped regime \cite{RuebhausenPRL1999,MaksimovACMP2010}. Together with the decrease of well-defined high-energy magnetic signal with doping in INS measurements \cite{WakimotoPRL2007,*LipscombePRL2007,*StockPRB2010}, this has cast doubt on the interpretation of the RIXS results \cite{TranquadaCondmat2013}.

The RIXS cross section in the cuprates is known to exhibit a non-trivial photon energy dependence \cite{LuPRL2005,LuPRB2006}, and the detection of magnetic excitations is greatly enhanced by a resonant process that creates an intermediate state with strong spin-orbit coupling \cite{AmentPRL2009,*HaverkortPRL2010}. Similarly, it has been known that the ERS detection of two-magnon excitations is assisted by resonant Raman processes \cite{ShastryIJMPB1991,DevereauxRMP2007}. For undoped systems, a strong enhancement of the signal is found for incident photon energies in the 2.4-3.0 eV range \cite{BlumbergPRB1994,CooperPRB1993,YoshidaPRB1992,RuebhausenPRL1999}, and it has been common experimental practice to use fixed laser energies suitable for the undoped systems to perform measurements at finite doping \cite{RuebhausenPRB1997,SugaiPRB2003}. A caveat is that the doping dependence of the ERS signal measured in this way may reflect variations not only in the magnetism but also in the resonance condition \cite{CooperPRB1993}.

We have determined the origin of the discrepancy between the RIXS and ERS results by performing ERS measurements in the model high-temperature superconductor HgBa$_2$CuO$_{4+\delta}$ (Hg1201).  We used two distinct incident photon energies for our ERS measurements and performed complementary ellipsometry measurements on the same samples, in order to monitor any possible change in the resonance condition. At low doping, the ERS two-magnon signal is resonantly enhanced with the higher incident photon energy, but as the hole concentration increases beyond $p\approx0.10$, the enhancement condition changes and favors the lower incident photon energy. This change coincides with a rearrangement of inter-band transitions in the 1-3 eV range observed by ellipsometry. Our clear observation of the two-magnon signal at doping levels as high as $p\approx0.19$, albeit only under a resonance condition that is different from that for undoped and lightly doped cuprates, demonstrates that the amplitude of the ERS signal is predominantly affected by the resonant process, and that short-range high-energy AF fluctuations do exist up to rather high doping. Furthermore, we find that the ERS $B_{1g}$ superconducting energy gap reaches its maximum near the same doping where the crossover between different resonance conditions occurs.

Our measurements were performed on eight single crystals of Hg1201 grown by the self-flux method \cite{ZhaoAdvMater2006} and post-growth annealed to achieve different doping \cite{BarisicPRB2008}. The samples are denoted by ``UD'' (underdoped) or ``OV'' (overdoped) followed by their critical temperature ($T_\mathrm{c}$) values in kelvin. Nominal hole concentrations are calculated from $T_\mathrm{c}$ according to an empirical formula \cite{TallonPRB1995}. The ERS measurements were performed in a quasi-backscattering geometry on freshly prepared surfaces parallel to the $ab$-plane using a Jobin Yvon LabRam spectrometer. The ellipsometry measurements were performed on a Woollam VASE spectrometer. Details about the measurement conditions can be found in the Supplemental Material.

\begin{figure}
\includegraphics[width=3.375in]{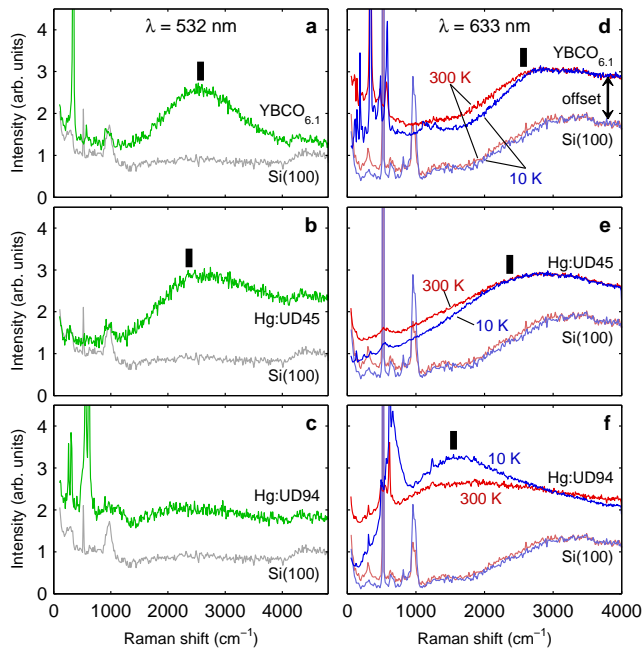}
\caption{\label{fig:one}
Raw Raman spectra for YBa$_2$Cu$_3$O$_{6.1}$ (a,d), Hg1201 UD45 (b,e) and UD94 (c,f) obtained with 532 nm (a-c, measured at $T=300$ K) and 633 nm lasers (d-f) in the $B_{1g}$ scattering geometry. Data obtained under the same condition on Si(100) surface are displayed for comparison (see text).
}
\end{figure}

Figure~\ref{fig:one} displays our Raman $B_{1g}$ spectra for a heavily (UD45) and a slightly underdoped (UD94) Hg1201 sample, along with data measured on an AF insulating YBa$_2$Cu$_3$O$_{6.1}$ (YBCO$_{6.1}$) sample. Spectra measured on a silicon (100) surface, which gives no ERS or fluorescence signal in the displayed range, are used for instrument correction (see Supplemental Material), and we only consider features that are absent from the Si spectra as genuine ERS signal. In YBCO$_{6.1}$, the two-magnon peak is observed at about 2600 cm$^{-1}$ with 532 nm incident photons, but not with 633 nm incident photons. In the latter ``off-resonance'' condition, the ERS signal shows a depletion at low temperature below 2600 cm$^{-1}$, which is the same as the peak position seen with the ``on-resonance'' condition. The Hg1201 UD45 sample looks very similar to YBCO$_{6.1}$, which is expected since YBCO at doping comparable to UD45 is not too different from the parent compound \cite{SugaiPRB2003}. However, a very different behavior is observed for UD94: the two-magnon peak can be observed with 633 nm incident photons already at $T=300$ K and its amplitude is enhanced at $T=10$ K \cite{LiPRL2012}, but no peak is observed with 532 nm incident photons. This can be best seen by comparing the data measured with the same incident photon energy between UD45 and UD94.

\begin{figure}
\includegraphics[width=3.375in]{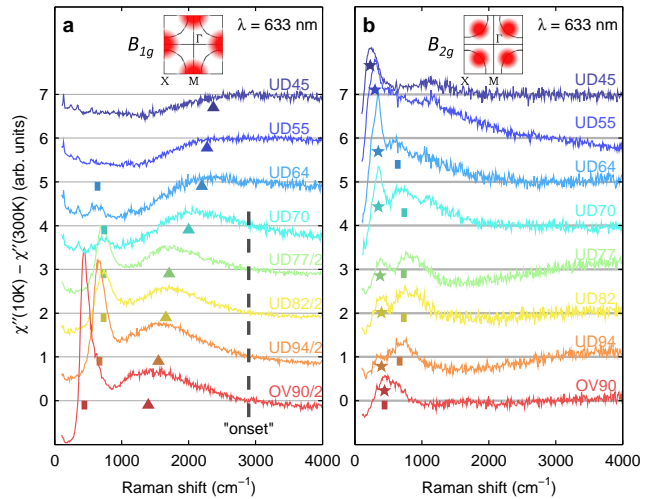}
\caption{\label{fig:two}
Difference between Raman susceptibility at 10 K and 300 K measured with 633 nm incident photons in the $B_{1g}$ (a) and $B_{2g}$ (b) geometries, offset for clarity. Characteristic energies are marked by rectangles ($B_{1g}$ peak, reproduced in panel b), stars ($B_{2g}$ peak), triangles ($B_{1g}$ two-magnon), and dashed line ($B_{1g}$ energy below which an enhancement of the two-magnon signal is observed at low temperature).
}
\end{figure}

Figure~\ref{fig:two}a shows this in a more systematic way by displaying the difference between $B_{1g}$ spectra taken at 10 K and 300 K with 633 nm incident photons for all our samples. A crossover from depletion to enhancement at low temperature is found near the doping of UD70. The enhancement becomes very prominent at the highest four doping levels, where the displayed spectra have been divided by two. The absence of similar effects in the $B_{2g}$ data (Fig.~\ref{fig:two}b) shows that the signal exclusively belongs to the $B_{1g}$ irreducible representation of the $D_{4h}$ point group, consistent with our interpretation that the signal evolves from two-magnon excitations in the undoped AF insulators. The characteristic energy determined from the position of either the depletion onset at low doping or the maximal enhancement at high doping (triangles in Fig.~\ref{fig:two}a), where the correspondence is supported by the on- and off-resonance measurements in Fig.~\ref{fig:one}, decreases with increasing doping as has been reported \cite{SugaiPRB2003,RuebhausenPRB1997}. In contrast, the highest energy at which an enhancement is found (dashed line in Fig.~\ref{fig:two}a) is independent of doping within our experimental accuracy. This constant energy might be connected to the doping-independent dispersion of paramagnons observed by RIXS near the magnetic zone boundary \cite{LeTaconNatPhys2011,LeTaconCondmat2013,*Deancondmat2013}. The strength and energy of the maximal enhancement (same as the peak-maximum position at 10 K) in our data might be further affected by the doping-dependent resonance condition, magnon-magnon interactions, and the size of the pseudogap. It therefore might not correspond to any observable feature in the single-magnon spectrum measured with RIXS.

\begin{figure}
\includegraphics[width=3.375in]{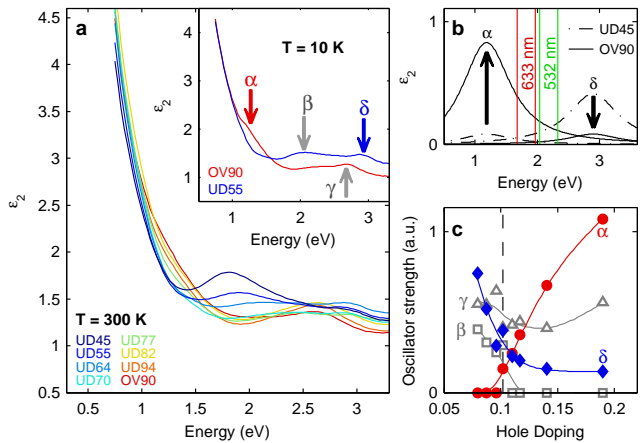}
\caption{\label{fig:three}
(a) Imaginary part of dielectric constant measured at 300 K. Inset: measurements at 10 K, where individual transitions are better seen due to reduced thermal broadening (see Supplemental Material for the full temperature dependence). (b) Schematic of the $\alpha$ and $\delta$ transitions relative to our ERS measurement conditions. The two vertical lines for each condition indicate the energies of the incident and scattered photons (after creating the two-magnon excitations). (c) Fit oscillator strength of the transitions labeled in (a).
}
\end{figure}

Figure~\ref{fig:three}a displays the imaginary part of the dielectric constant measured by ellipsometry on the same samples at 300 K.  In addition to a Drude response, the evolution with doping in the 0.8-3.2 eV range can be described by four Lorentzian oscillators modeling inter-band transitions, as labeled in the inset. The $\alpha$ and $\delta$ transitions are slightly below and above our ERS incident photon energies, respectively, so $\alpha$ favors 633 nm incident photons whereas $\delta$ favors 532 nm incident photons (Fig.~\ref{fig:three}b). $\beta$ is close to both our incident photon energies and thus cannot differentiate them, and $\gamma$ does not exhibit systematic doping dependence. The oscillator strengths of these transitions are obtained by simultaneously fitting the real and imaginary parts of the dielectric function (Supplemental Material) and summarized in Fig.~\ref{fig:three}c. Indeed, the $\delta$ transition is prominent at low doping while $\alpha$ is prominent at high doping, showing a crossover behavior near UD70 ($p\approx0.10$). This precisely corresponds to our observation of the ERS two-magnon signal with 532 nm and 633 nm incident photons at low and high doping, respectively. Our combined ERS and ellipsometry data show that short-range high-energy AF fluctuations are present in Hg1201 at least up to $p=0.19$, and that they can be observed by ERS if the doping-dependent resonance condition is satisfied. Hence there is no essential discrepancy between the ERS and RIXS results concerning the presence of short-range AF fluctuations deep in the superconducting regime, all the way to the overdoped side of the phase diagram.

The spectra in Fig.~\ref{fig:two} also contain low-energy ERS signals associated with superconductivity. They are marked by the rectangular and star symbols for the $B_{1g}$ and $B_{2g}$ geometries, which selectively probe electronic transitions in the antinodal and nodal regions of the Brillouin zone \cite{DevereauxRMP2007}, respectively. The data are in good overall agreement with reported ERS results for Hg1201 and other cuprates: the $B_{1g}$ energy decreases as optimal doping is approached from below while the $B_{2g}$ energy increases \cite{LeTaconNatPhys2006,BlancPRB2010,MunnikesPRB2011}, and the intensity of the $B_{1g}$ peak increases rapidly with overdoping \cite{MunnikesPRB2011}. Here we focus on the underdoped side and make a few observations:

(1) The decrease and eventual disappearance of the $B_{1g}$ peak with underdoping resembles those of the two-magnon peak. This suggests that the two features are enhanced by similar resonant effects.

(2) The $B_{1g}$ energy is also visible in the $B_{2g}$ data as indicated by rectangular symbols in Fig.~\ref{fig:two}b. While we do not know the exact reason for this, we can rule out polarization leakage which we estimate to be less than $3\%$ based on phonon intensities observed in the different geometries. Our data do not contradict previous reports \cite{LeTaconNatPhys2006,GallaisPRB2005,BlancPRB2010,MunnikesPRB2011}, where the use of different incident photon energies and/or a smaller separation between the $B_{1g}$ and $B_{2g}$ energies might have prevented a similar observation. At low doping, the $B_{2g}$ signal exhibits a long tail extending above 1000 cm$^{-1}$. This component of signal persists above $T_\mathrm{c}$ (not shown) and might be a signature of a pseudogap recently proposed to have an $s$-wave form \cite{SakaiCondmat2012}. The $B_{2g}$ double-peak structure resembles a recent theoretical proposal of Higgs-like excitations \cite{BarlasCondmat2013}.

(3) We observe the $B_{1g}$ peak down to an unprecedentedly low doping level of $p\approx0.09$ (UD64), and see a clear decrease of its energy below $p\approx0.10$. 
A similar decrease of the $d$-wave superconducting gap has been observed by photoemission for Bi2212 below $p\approx0.08$ \cite{TanakaScience2006,*VishikPNAS2012}.

\begin{figure}
\includegraphics[width=2.6in]{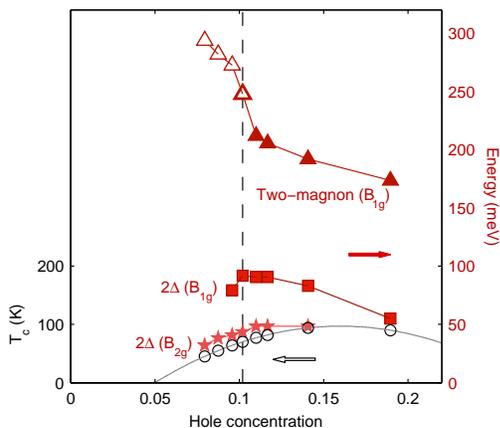}
\caption{\label{fig:four}
Summary of ERS characteristic energies. Same symbols are used as in Fig.~\ref{fig:two}.
}
\end{figure}

We summarize our observations in Fig.~\ref{fig:four}, in which the dashed line indicates where the rearrangement of inter-band transitions occurs. The two-magnon signal seen with 633 nm incident photons changes its temperature dependence from depletion (empty triangles) to enhancement (filled triangles) at low temperature upon crossing this line, where its characteristic energy exhibits a sudden decrease. Intriguingly, we find that the $B_{1g}$ superconducting gap exhibits a maximum near the same doping.

Our result sheds light on the nature of the relevant high-energy electronic states.  At low doping, the resonant Raman process has been a subject of considerable theoretical investigation \cite{ShastryIJMPB1991,ChubukovPRL1995,*ChubukovPRB1995,*HanamuraPRB2000,*TohyamaPRL2002,*KupcicJRS2011}, and it generally requires the absorption of a photon that is energetic enough to overcome the effective Hubbard repulsion, or the charge-transfer gap. It can be seen from Fig.~\ref{fig:three}a, however, that the $\alpha$ transition does not evolve continuously out of any of the other transitions: $\beta$ is the only transition that clearly moves with doping but it moves away from $\alpha$. Therefore, it seems that the $\alpha$ transition involves a new band that develops at high doping, inside the charge-transfer gap, which appears to be present in Hg1201 up to at least optimal doping \cite{LuPRL2005}. Similar development of a new inter-band transition in the 0.8-1.5 eV range at high doping has been found in YBCO (ref.~\onlinecite{CooperPRB1993}), La$_{2-x}$Sr$_x$CuO$_4$ (ref.~\onlinecite{UchidaPRB1991}), and Bi$_2$Sr$_2$Ca$_{0.92}$Y$_{0.08}$Cu$_2$O$_{8+\delta}$ (ref.~\onlinecite{GiannettiNatComm2011}), and it has been attributed to a transition from a localized polaronic state inside the charge-transfer gap to an extended state above the gap \cite{LorenzanaPRL1993}. Other possibilities, such as a breakdown of the Zhang-Rice singlet approximation at finite doping \cite{SchneiderPRB2005,*PeetsPRL2009}, doped holes not entering the planar orbitals \cite{ChenPRL1992,*MerzPRL1998}, or splitting of the charge-transfer peak with the suppression of AF correlation \cite{SedrakyanPRB2010}, cannot be ruled out at this time. Band-structure calculations for undoped Hg1201 indicate the presence of a Hg-O band not far from the Fermi level which may further evolve with doping \cite{SakakibaraPRB2012,*DasPRB2012}, but this scenario would have difficulty explaining the resonance effect on two-magnon excitations given the large spatial separation between the Cu-O and Hg layers. First-principle calculations for doped Hg1201 are technically challenging due to the complex dopant oxygen positions, which have not yet been fully determined \cite{IzquierdoJPhysChemSolids2011,*ChabotCouturePreprint}.

To conclude, we have identified a dramatic change in the Raman resonance condition with doping in Hg1201, both by direct observation of the two-magnon ERS signal at high doping under an unusual condition and by ellipsometry observation of a rearrangement of inter-band transitions near $p=0.10$. Our data are consistent with the presence of short-range high-energy AF fluctuations even in the overdoped regime, and they allow us to reconcile the discrepancy between existing ERS and RIXS results. Further research is needed to understand the exact nature of the high-energy electronic states involved in the Raman resonance, which appear to affect the size of the superconducting gap.

\begin{acknowledgments}
We wish to thank P. Bourges, A. Chubukov, M. Civelli, T. Devereaux, T. Tohyama, C.M. Varma, Z.-Y. Weng, and A. Yaresko for stimulating discussions and Armin Schultz for technical assistance. Y.L. acknowledges the Alexander von Humboldt foundation and start-up support from Peking University. Crystal growth and characterization work at the University of Minnesota is supported by the Department of Energy, Office of Basic Energy Sciences.
\end{acknowledgments}

\bibliography{Raman_Ellipsometry_Hg1201_v2}

\begin{thebibliography}{55}%
\makeatletter
\providecommand \@ifxundefined [1]{%
 \@ifx{#1\undefined}
}%
\providecommand \@ifnum [1]{%
 \ifnum #1\expandafter \@firstoftwo
 \else \expandafter \@secondoftwo
 \fi
}%
\providecommand \@ifx [1]{%
 \ifx #1\expandafter \@firstoftwo
 \else \expandafter \@secondoftwo
 \fi
}%
\providecommand \natexlab [1]{#1}%
\providecommand \enquote  [1]{``#1''}%
\providecommand \bibnamefont  [1]{#1}%
\providecommand \bibfnamefont [1]{#1}%
\providecommand \citenamefont [1]{#1}%
\providecommand \href@noop [0]{\@secondoftwo}%
\providecommand \href [0]{\begingroup \@sanitize@url \@href}%
\providecommand \@href[1]{\@@startlink{#1}\@@href}%
\providecommand \@@href[1]{\endgroup#1\@@endlink}%
\providecommand \@sanitize@url [0]{\catcode `\\12\catcode `\$12\catcode
  `\&12\catcode `\#12\catcode `\^12\catcode `\_12\catcode `\%12\relax}%
\providecommand \@@startlink[1]{}%
\providecommand \@@endlink[0]{}%
\providecommand \url  [0]{\begingroup\@sanitize@url \@url }%
\providecommand \@url [1]{\endgroup\@href {#1}{\urlprefix }}%
\providecommand \urlprefix  [0]{URL }%
\providecommand \Eprint [0]{\href }%
\providecommand \doibase [0]{http://dx.doi.org/}%
\providecommand \selectlanguage [0]{\@gobble}%
\providecommand \bibinfo  [0]{\@secondoftwo}%
\providecommand \bibfield  [0]{\@secondoftwo}%
\providecommand \translation [1]{[#1]}%
\providecommand \BibitemOpen [0]{}%
\providecommand \bibitemStop [0]{}%
\providecommand \bibitemNoStop [0]{.\EOS\space}%
\providecommand \EOS [0]{\spacefactor3000\relax}%
\providecommand \BibitemShut  [1]{\csname bibitem#1\endcsname}%
\let\auto@bib@innerbib\@empty
\bibitem [{\citenamefont {Scalapino}(1995)}]{ScalapinoPhysRep1995}%
  \BibitemOpen
  \bibfield  {author} {\bibinfo {author} {\bibfnamefont {D.~J.}\ \bibnamefont
  {Scalapino}},\ }\href {\doibase 10.1016/0370-1573(94)00086-I} {\bibfield
  {journal} {\bibinfo  {journal} {Phys. Rep.}\ }\textbf {\bibinfo {volume}
  {250}},\ \bibinfo {pages} {329} (\bibinfo {year} {1995})}\BibitemShut
  {NoStop}%
\bibitem [{\citenamefont {Abanov}\ \emph {et~al.}(2003)\citenamefont {Abanov},
  \citenamefont {Chubukov},\ and\ \citenamefont
  {Schmalian}}]{AbanovAdvPhys2003}%
  \BibitemOpen
  \bibfield  {author} {\bibinfo {author} {\bibfnamefont {A.}~\bibnamefont
  {Abanov}}, \bibinfo {author} {\bibfnamefont {A.~V.}\ \bibnamefont
  {Chubukov}}, \ and\ \bibinfo {author} {\bibfnamefont {J.}~\bibnamefont
  {Schmalian}},\ }\href {\doibase 10.1080/0001873021000057123} {\bibfield
  {journal} {\bibinfo  {journal} {Adv. Phys.}\ }\textbf {\bibinfo {volume}
  {52}},\ \bibinfo {pages} {119} (\bibinfo {year} {2003})}\BibitemShut
  {NoStop}%
\bibitem [{\citenamefont {Moriya}\ and\ \citenamefont
  {Ueda}(2003)}]{MoriyaRepProgPhys2003}%
  \BibitemOpen
  \bibfield  {author} {\bibinfo {author} {\bibfnamefont {T.}~\bibnamefont
  {Moriya}}\ and\ \bibinfo {author} {\bibfnamefont {K.}~\bibnamefont {Ueda}},\
  }\href {http://stacks.iop.org/0034-4885/66/i=8/a=202} {\bibfield  {journal}
  {\bibinfo  {journal} {Rep. Prog. Phys.}\ }\textbf {\bibinfo {volume} {66}},\
  \bibinfo {pages} {1299} (\bibinfo {year} {2003})}\BibitemShut {NoStop}%
\bibitem [{\citenamefont {Le~Tacon}\ \emph {et~al.}(2011)\citenamefont
  {Le~Tacon}, \citenamefont {Ghiringhelli}, \citenamefont {Chaloupka},
  \citenamefont {Moretti~Sala}, \citenamefont {Hinkov}, \citenamefont
  {Haverkort}, \citenamefont {Minola}, \citenamefont {Bakr}, \citenamefont
  {Zhou}, \citenamefont {Blanco-Canosa}, \citenamefont {Monney}, \citenamefont
  {Song}, \citenamefont {Sun}, \citenamefont {Lin}, \citenamefont {De~Luca},
  \citenamefont {Salluzzo}, \citenamefont {Khaliullin}, \citenamefont
  {Schmitt}, \citenamefont {Braicovich},\ and\ \citenamefont
  {Keimer}}]{LeTaconNatPhys2011}%
  \BibitemOpen
  \bibfield  {author} {\bibinfo {author} {\bibfnamefont {M.}~\bibnamefont
  {Le~Tacon}}, \bibinfo {author} {\bibfnamefont {G.}~\bibnamefont
  {Ghiringhelli}}, \bibinfo {author} {\bibfnamefont {J.}~\bibnamefont
  {Chaloupka}}, \bibinfo {author} {\bibfnamefont {M.}~\bibnamefont
  {Moretti~Sala}}, \bibinfo {author} {\bibfnamefont {V.}~\bibnamefont
  {Hinkov}}, \bibinfo {author} {\bibfnamefont {M.~W.}\ \bibnamefont
  {Haverkort}}, \bibinfo {author} {\bibfnamefont {M.}~\bibnamefont {Minola}},
  \bibinfo {author} {\bibfnamefont {M.}~\bibnamefont {Bakr}}, \bibinfo {author}
  {\bibfnamefont {K.~J.}\ \bibnamefont {Zhou}}, \bibinfo {author}
  {\bibfnamefont {S.}~\bibnamefont {Blanco-Canosa}}, \bibinfo {author}
  {\bibfnamefont {C.}~\bibnamefont {Monney}}, \bibinfo {author} {\bibfnamefont
  {Y.~T.}\ \bibnamefont {Song}}, \bibinfo {author} {\bibfnamefont {G.~L.}\
  \bibnamefont {Sun}}, \bibinfo {author} {\bibfnamefont {C.~T.}\ \bibnamefont
  {Lin}}, \bibinfo {author} {\bibfnamefont {G.~M.}\ \bibnamefont {De~Luca}},
  \bibinfo {author} {\bibfnamefont {M.}~\bibnamefont {Salluzzo}}, \bibinfo
  {author} {\bibfnamefont {G.}~\bibnamefont {Khaliullin}}, \bibinfo {author}
  {\bibfnamefont {T.}~\bibnamefont {Schmitt}}, \bibinfo {author} {\bibfnamefont
  {L.}~\bibnamefont {Braicovich}}, \ and\ \bibinfo {author} {\bibfnamefont
  {B.}~\bibnamefont {Keimer}},\ }\href@noop {} {\bibfield  {journal} {\bibinfo
  {journal} {Nature Phys.}\ }\textbf {\bibinfo {volume} {7}},\ \bibinfo {pages}
  {725} (\bibinfo {year} {2011})}\BibitemShut {NoStop}%
\bibitem [{\citenamefont {Li}\ \emph {et~al.}(2012)\citenamefont {Li},
  \citenamefont {Le~Tacon}, \citenamefont {Bakr}, \citenamefont {Terrade},
  \citenamefont {Manske}, \citenamefont {Hackl}, \citenamefont {Ji},
  \citenamefont {Chan}, \citenamefont {Bari\ifmmode \check{s}\else
  \v{s}\fi{}i\ifmmode~\acute{c}\else \'{c}\fi{}}, \citenamefont {Zhao},
  \citenamefont {Greven},\ and\ \citenamefont {Keimer}}]{LiPRL2012}%
  \BibitemOpen
  \bibfield  {author} {\bibinfo {author} {\bibfnamefont {Y.}~\bibnamefont
  {Li}}, \bibinfo {author} {\bibfnamefont {M.}~\bibnamefont {Le~Tacon}},
  \bibinfo {author} {\bibfnamefont {M.}~\bibnamefont {Bakr}}, \bibinfo {author}
  {\bibfnamefont {D.}~\bibnamefont {Terrade}}, \bibinfo {author} {\bibfnamefont
  {D.}~\bibnamefont {Manske}}, \bibinfo {author} {\bibfnamefont
  {R.}~\bibnamefont {Hackl}}, \bibinfo {author} {\bibfnamefont
  {L.}~\bibnamefont {Ji}}, \bibinfo {author} {\bibfnamefont {M.~K.}\
  \bibnamefont {Chan}}, \bibinfo {author} {\bibfnamefont {N.}~\bibnamefont
  {Bari\ifmmode \check{s}\else \v{s}\fi{}i\ifmmode~\acute{c}\else \'{c}\fi{}}},
  \bibinfo {author} {\bibfnamefont {X.}~\bibnamefont {Zhao}}, \bibinfo {author}
  {\bibfnamefont {M.}~\bibnamefont {Greven}}, \ and\ \bibinfo {author}
  {\bibfnamefont {B.}~\bibnamefont {Keimer}},\ }\href {\doibase
  10.1103/PhysRevLett.108.227003} {\bibfield  {journal} {\bibinfo  {journal}
  {Phys. Rev. Lett.}\ }\textbf {\bibinfo {volume} {108}},\ \bibinfo {pages}
  {227003} (\bibinfo {year} {2012})}\BibitemShut {NoStop}%
\bibitem [{\citenamefont {Braicovich}\ \emph {et~al.}(2010)\citenamefont
  {Braicovich}, \citenamefont {van~den Brink}, \citenamefont {Bisogni},
  \citenamefont {Sala}, \citenamefont {Ament}, \citenamefont {Brookes},
  \citenamefont {De~Luca}, \citenamefont {Salluzzo}, \citenamefont {Schmitt},
  \citenamefont {Strocov},\ and\ \citenamefont
  {Ghiringhelli}}]{BraicovichPRL2010}%
  \BibitemOpen
  \bibfield  {author} {\bibinfo {author} {\bibfnamefont {L.}~\bibnamefont
  {Braicovich}}, \bibinfo {author} {\bibfnamefont {J.}~\bibnamefont {van~den
  Brink}}, \bibinfo {author} {\bibfnamefont {V.}~\bibnamefont {Bisogni}},
  \bibinfo {author} {\bibfnamefont {M.~M.}\ \bibnamefont {Sala}}, \bibinfo
  {author} {\bibfnamefont {L.~J.~P.}\ \bibnamefont {Ament}}, \bibinfo {author}
  {\bibfnamefont {N.~B.}\ \bibnamefont {Brookes}}, \bibinfo {author}
  {\bibfnamefont {G.~M.}\ \bibnamefont {De~Luca}}, \bibinfo {author}
  {\bibfnamefont {M.}~\bibnamefont {Salluzzo}}, \bibinfo {author}
  {\bibfnamefont {T.}~\bibnamefont {Schmitt}}, \bibinfo {author} {\bibfnamefont
  {V.~N.}\ \bibnamefont {Strocov}}, \ and\ \bibinfo {author} {\bibfnamefont
  {G.}~\bibnamefont {Ghiringhelli}},\ }\href {\doibase
  10.1103/PhysRevLett.104.077002} {\bibfield  {journal} {\bibinfo  {journal}
  {Phys. Rev. Lett.}\ }\textbf {\bibinfo {volume} {104}},\ \bibinfo {pages}
  {077002} (\bibinfo {year} {2010})}\BibitemShut {NoStop}%
\bibitem [{\citenamefont {Lyons}\ \emph {et~al.}(1988)\citenamefont {Lyons},
  \citenamefont {Fleury}, \citenamefont {Remeika}, \citenamefont {Cooper},\
  and\ \citenamefont {Negran}}]{LyonsPRB1988}%
  \BibitemOpen
  \bibfield  {author} {\bibinfo {author} {\bibfnamefont {K.~B.}\ \bibnamefont
  {Lyons}}, \bibinfo {author} {\bibfnamefont {P.~A.}\ \bibnamefont {Fleury}},
  \bibinfo {author} {\bibfnamefont {J.~P.}\ \bibnamefont {Remeika}}, \bibinfo
  {author} {\bibfnamefont {A.~S.}\ \bibnamefont {Cooper}}, \ and\ \bibinfo
  {author} {\bibfnamefont {T.~J.}\ \bibnamefont {Negran}},\ }\href {\doibase
  10.1103/PhysRevB.37.2353} {\bibfield  {journal} {\bibinfo  {journal} {Phys.
  Rev. B}\ }\textbf {\bibinfo {volume} {37}},\ \bibinfo {pages} {2353}
  (\bibinfo {year} {1988})}\BibitemShut {NoStop}%
\bibitem [{\citenamefont {Coldea}\ \emph {et~al.}(2001)\citenamefont {Coldea},
  \citenamefont {Hayden}, \citenamefont {Aeppli}, \citenamefont {Perring},
  \citenamefont {Frost}, \citenamefont {Mason}, \citenamefont {Cheong},\ and\
  \citenamefont {Fisk}}]{ColdeaPRL2001}%
  \BibitemOpen
  \bibfield  {author} {\bibinfo {author} {\bibfnamefont {R.}~\bibnamefont
  {Coldea}}, \bibinfo {author} {\bibfnamefont {S.~M.}\ \bibnamefont {Hayden}},
  \bibinfo {author} {\bibfnamefont {G.}~\bibnamefont {Aeppli}}, \bibinfo
  {author} {\bibfnamefont {T.~G.}\ \bibnamefont {Perring}}, \bibinfo {author}
  {\bibfnamefont {C.~D.}\ \bibnamefont {Frost}}, \bibinfo {author}
  {\bibfnamefont {T.~E.}\ \bibnamefont {Mason}}, \bibinfo {author}
  {\bibfnamefont {S.-W.}\ \bibnamefont {Cheong}}, \ and\ \bibinfo {author}
  {\bibfnamefont {Z.}~\bibnamefont {Fisk}},\ }\href {\doibase
  10.1103/PhysRevLett.86.5377} {\bibfield  {journal} {\bibinfo  {journal}
  {Phys. Rev. Lett.}\ }\textbf {\bibinfo {volume} {86}},\ \bibinfo {pages}
  {5377} (\bibinfo {year} {2001})}\BibitemShut {NoStop}%
\bibitem [{\citenamefont {Le~Tacon}\ \emph {et~al.}()\citenamefont {Le~Tacon},
  \citenamefont {Minola}, \citenamefont {Peets}, \citenamefont {Moretti~Sala},
  \citenamefont {Blanco-Canosa}, \citenamefont {Hinkov}, \citenamefont {Liang},
  \citenamefont {Bonn}, \citenamefont {Hardy}, \citenamefont {Lin},
  \citenamefont {Schmitt}, \citenamefont {Braicovich}, \citenamefont
  {Ghiringhelli},\ and\ \citenamefont {Keimer}}]{LeTaconCondmat2013}%
  \BibitemOpen
  \bibfield  {author} {\bibinfo {author} {\bibfnamefont {M.}~\bibnamefont
  {Le~Tacon}}, \bibinfo {author} {\bibfnamefont {M.}~\bibnamefont {Minola}},
  \bibinfo {author} {\bibfnamefont {D.~C.}\ \bibnamefont {Peets}}, \bibinfo
  {author} {\bibfnamefont {M.}~\bibnamefont {Moretti~Sala}}, \bibinfo {author}
  {\bibfnamefont {S.}~\bibnamefont {Blanco-Canosa}}, \bibinfo {author}
  {\bibfnamefont {V.}~\bibnamefont {Hinkov}}, \bibinfo {author} {\bibfnamefont
  {R.}~\bibnamefont {Liang}}, \bibinfo {author} {\bibfnamefont {D.~A.}\
  \bibnamefont {Bonn}}, \bibinfo {author} {\bibfnamefont {W.~N.}\ \bibnamefont
  {Hardy}}, \bibinfo {author} {\bibfnamefont {C.~T.}\ \bibnamefont {Lin}},
  \bibinfo {author} {\bibfnamefont {T.}~\bibnamefont {Schmitt}}, \bibinfo
  {author} {\bibfnamefont {L.}~\bibnamefont {Braicovich}}, \bibinfo {author}
  {\bibfnamefont {G.}~\bibnamefont {Ghiringhelli}}, \ and\ \bibinfo {author}
  {\bibfnamefont {B.}~\bibnamefont {Keimer}},\ }\href@noop {} {\enquote
  {\bibinfo {title} {Dispersive spin excitations in highly overdoped cuprates
  revealed by resonant inelastic x-ray scattering},}\ }\bibinfo {note}
  {ArXiv:1303.3947}\BibitemShut {NoStop}%
\bibitem [{\citenamefont {Dean}\ \emph {et~al.}()\citenamefont {Dean},
  \citenamefont {Dellea}, \citenamefont {Springell}, \citenamefont
  {Yakhou-Harris}, \citenamefont {Kummer}, \citenamefont {Brooks},
  \citenamefont {Liu}, \citenamefont {Sun}, \citenamefont {Strle},
  \citenamefont {Schmitt}, \citenamefont {Braicovich}, \citenamefont
  {Ghiringhelli}, \citenamefont {Bozovic},\ and\ \citenamefont
  {Hill}}]{Deancondmat2013}%
  \BibitemOpen
  \bibfield  {author} {\bibinfo {author} {\bibfnamefont {M.~P.~M.}\
  \bibnamefont {Dean}}, \bibinfo {author} {\bibfnamefont {G.}~\bibnamefont
  {Dellea}}, \bibinfo {author} {\bibfnamefont {R.~S.}\ \bibnamefont
  {Springell}}, \bibinfo {author} {\bibfnamefont {F.}~\bibnamefont
  {Yakhou-Harris}}, \bibinfo {author} {\bibfnamefont {K.}~\bibnamefont
  {Kummer}}, \bibinfo {author} {\bibfnamefont {N.~B.}\ \bibnamefont {Brooks}},
  \bibinfo {author} {\bibfnamefont {X.}~\bibnamefont {Liu}}, \bibinfo {author}
  {\bibfnamefont {Y.-J.}\ \bibnamefont {Sun}}, \bibinfo {author} {\bibfnamefont
  {J.}~\bibnamefont {Strle}}, \bibinfo {author} {\bibfnamefont
  {T.}~\bibnamefont {Schmitt}}, \bibinfo {author} {\bibfnamefont
  {L.}~\bibnamefont {Braicovich}}, \bibinfo {author} {\bibfnamefont
  {G.}~\bibnamefont {Ghiringhelli}}, \bibinfo {author} {\bibfnamefont
  {I.}~\bibnamefont {Bozovic}}, \ and\ \bibinfo {author} {\bibfnamefont
  {J.~P.}\ \bibnamefont {Hill}},\ }\href@noop {} {\enquote {\bibinfo {title}
  {Persistence of magnetic excitations in {La$_{2-x}$Sr$_x$CuO$_4$} from the
  underdoped insulator to the heavily overdoped non-superconducting metal},}\
  }\bibinfo {note} {ArXiv:1303.5359}\BibitemShut {NoStop}%
\bibitem [{\citenamefont {Cooper}\ \emph {et~al.}(1993)\citenamefont {Cooper},
  \citenamefont {Reznik}, \citenamefont {Kotz}, \citenamefont {Karlow},
  \citenamefont {Liu}, \citenamefont {Klein}, \citenamefont {Lee},
  \citenamefont {Giapintzakis}, \citenamefont {Ginsberg}, \citenamefont
  {Veal},\ and\ \citenamefont {Paulikas}}]{CooperPRB1993}%
  \BibitemOpen
  \bibfield  {author} {\bibinfo {author} {\bibfnamefont {S.~L.}\ \bibnamefont
  {Cooper}}, \bibinfo {author} {\bibfnamefont {D.}~\bibnamefont {Reznik}},
  \bibinfo {author} {\bibfnamefont {A.}~\bibnamefont {Kotz}}, \bibinfo {author}
  {\bibfnamefont {M.~A.}\ \bibnamefont {Karlow}}, \bibinfo {author}
  {\bibfnamefont {R.}~\bibnamefont {Liu}}, \bibinfo {author} {\bibfnamefont
  {M.~V.}\ \bibnamefont {Klein}}, \bibinfo {author} {\bibfnamefont {W.~C.}\
  \bibnamefont {Lee}}, \bibinfo {author} {\bibfnamefont {J.}~\bibnamefont
  {Giapintzakis}}, \bibinfo {author} {\bibfnamefont {D.~M.}\ \bibnamefont
  {Ginsberg}}, \bibinfo {author} {\bibfnamefont {B.~W.}\ \bibnamefont {Veal}},
  \ and\ \bibinfo {author} {\bibfnamefont {A.~P.}\ \bibnamefont {Paulikas}},\
  }\href {\doibase 10.1103/PhysRevB.47.8233} {\bibfield  {journal} {\bibinfo
  {journal} {Phys. Rev. B}\ }\textbf {\bibinfo {volume} {47}},\ \bibinfo
  {pages} {8233} (\bibinfo {year} {1993})}\BibitemShut {NoStop}%
\bibitem [{\citenamefont {Blumberg}\ \emph {et~al.}(1994)\citenamefont
  {Blumberg}, \citenamefont {Liu}, \citenamefont {Klein}, \citenamefont {Lee},
  \citenamefont {Ginsberg}, \citenamefont {Gu}, \citenamefont {Veal},\ and\
  \citenamefont {Dabrowski}}]{BlumbergPRB1994}%
  \BibitemOpen
  \bibfield  {author} {\bibinfo {author} {\bibfnamefont {G.}~\bibnamefont
  {Blumberg}}, \bibinfo {author} {\bibfnamefont {R.}~\bibnamefont {Liu}},
  \bibinfo {author} {\bibfnamefont {M.~V.}\ \bibnamefont {Klein}}, \bibinfo
  {author} {\bibfnamefont {W.~C.}\ \bibnamefont {Lee}}, \bibinfo {author}
  {\bibfnamefont {D.~M.}\ \bibnamefont {Ginsberg}}, \bibinfo {author}
  {\bibfnamefont {C.}~\bibnamefont {Gu}}, \bibinfo {author} {\bibfnamefont
  {B.~W.}\ \bibnamefont {Veal}}, \ and\ \bibinfo {author} {\bibfnamefont
  {B.}~\bibnamefont {Dabrowski}},\ }\href {\doibase 10.1103/PhysRevB.49.13295}
  {\bibfield  {journal} {\bibinfo  {journal} {Phys. Rev. B}\ }\textbf {\bibinfo
  {volume} {49}},\ \bibinfo {pages} {13295} (\bibinfo {year}
  {1994})}\BibitemShut {NoStop}%
\bibitem [{\citenamefont {R\"ubhausen}\ \emph {et~al.}(1997)\citenamefont
  {R\"ubhausen}, \citenamefont {Rieck}, \citenamefont {Dieckmann},
  \citenamefont {Subke}, \citenamefont {Bock},\ and\ \citenamefont
  {Merkt}}]{RuebhausenPRB1997}%
  \BibitemOpen
  \bibfield  {author} {\bibinfo {author} {\bibfnamefont {M.}~\bibnamefont
  {R\"ubhausen}}, \bibinfo {author} {\bibfnamefont {C.~T.}\ \bibnamefont
  {Rieck}}, \bibinfo {author} {\bibfnamefont {N.}~\bibnamefont {Dieckmann}},
  \bibinfo {author} {\bibfnamefont {K.-O.}\ \bibnamefont {Subke}}, \bibinfo
  {author} {\bibfnamefont {A.}~\bibnamefont {Bock}}, \ and\ \bibinfo {author}
  {\bibfnamefont {U.}~\bibnamefont {Merkt}},\ }\href {\doibase
  10.1103/PhysRevB.56.14797} {\bibfield  {journal} {\bibinfo  {journal} {Phys.
  Rev. B}\ }\textbf {\bibinfo {volume} {56}},\ \bibinfo {pages} {14797}
  (\bibinfo {year} {1997})}\BibitemShut {NoStop}%
\bibitem [{\citenamefont {Sugai}\ \emph {et~al.}(2003)\citenamefont {Sugai},
  \citenamefont {Suzuki}, \citenamefont {Takayanagi}, \citenamefont
  {Hosokawa},\ and\ \citenamefont {Hayamizu}}]{SugaiPRB2003}%
  \BibitemOpen
  \bibfield  {author} {\bibinfo {author} {\bibfnamefont {S.}~\bibnamefont
  {Sugai}}, \bibinfo {author} {\bibfnamefont {H.}~\bibnamefont {Suzuki}},
  \bibinfo {author} {\bibfnamefont {Y.}~\bibnamefont {Takayanagi}}, \bibinfo
  {author} {\bibfnamefont {T.}~\bibnamefont {Hosokawa}}, \ and\ \bibinfo
  {author} {\bibfnamefont {N.}~\bibnamefont {Hayamizu}},\ }\href {\doibase
  10.1103/PhysRevB.68.184504} {\bibfield  {journal} {\bibinfo  {journal} {Phys.
  Rev. B}\ }\textbf {\bibinfo {volume} {68}},\ \bibinfo {pages} {184504}
  (\bibinfo {year} {2003})}\BibitemShut {NoStop}%
\bibitem [{\citenamefont {R\"ubhausen}\ \emph {et~al.}(1999)\citenamefont
  {R\"ubhausen}, \citenamefont {Hammerstein}, \citenamefont {Bock},
  \citenamefont {Merkt}, \citenamefont {Rieck}, \citenamefont {Guptasarma},
  \citenamefont {Hinks},\ and\ \citenamefont {Klein}}]{RuebhausenPRL1999}%
  \BibitemOpen
  \bibfield  {author} {\bibinfo {author} {\bibfnamefont {M.}~\bibnamefont
  {R\"ubhausen}}, \bibinfo {author} {\bibfnamefont {O.~A.}\ \bibnamefont
  {Hammerstein}}, \bibinfo {author} {\bibfnamefont {A.}~\bibnamefont {Bock}},
  \bibinfo {author} {\bibfnamefont {U.}~\bibnamefont {Merkt}}, \bibinfo
  {author} {\bibfnamefont {C.~T.}\ \bibnamefont {Rieck}}, \bibinfo {author}
  {\bibfnamefont {P.}~\bibnamefont {Guptasarma}}, \bibinfo {author}
  {\bibfnamefont {D.~G.}\ \bibnamefont {Hinks}}, \ and\ \bibinfo {author}
  {\bibfnamefont {M.~V.}\ \bibnamefont {Klein}},\ }\href {\doibase
  10.1103/PhysRevLett.82.5349} {\bibfield  {journal} {\bibinfo  {journal}
  {Phys. Rev. Lett.}\ }\textbf {\bibinfo {volume} {82}},\ \bibinfo {pages}
  {5349} (\bibinfo {year} {1999})}\BibitemShut {NoStop}%
\bibitem [{\citenamefont {Maksimov}\ \emph {et~al.}(2010)\citenamefont
  {Maksimov}, \citenamefont {Kuli\'c},\ and\ \citenamefont
  {Dolgov}}]{MaksimovACMP2010}%
  \BibitemOpen
  \bibfield  {author} {\bibinfo {author} {\bibfnamefont {E.~G.}\ \bibnamefont
  {Maksimov}}, \bibinfo {author} {\bibfnamefont {M.~L.}\ \bibnamefont
  {Kuli\'c}}, \ and\ \bibinfo {author} {\bibfnamefont {O.~V.}\ \bibnamefont
  {Dolgov}},\ }\href@noop {} {\bibfield  {journal} {\bibinfo  {journal}
  {Advances in Condensed Matter Physics}\ }\textbf {\bibinfo {volume} {2010}},\
  \bibinfo {pages} {423725} (\bibinfo {year} {2010})}\BibitemShut {NoStop}%
\bibitem [{\citenamefont {Wakimoto}\ \emph {et~al.}(2007)\citenamefont
  {Wakimoto}, \citenamefont {Yamada}, \citenamefont {Tranquada}, \citenamefont
  {Frost}, \citenamefont {Birgeneau},\ and\ \citenamefont
  {Zhang}}]{WakimotoPRL2007}%
  \BibitemOpen
  \bibfield  {author} {\bibinfo {author} {\bibfnamefont {S.}~\bibnamefont
  {Wakimoto}}, \bibinfo {author} {\bibfnamefont {K.}~\bibnamefont {Yamada}},
  \bibinfo {author} {\bibfnamefont {J.~M.}\ \bibnamefont {Tranquada}}, \bibinfo
  {author} {\bibfnamefont {C.~D.}\ \bibnamefont {Frost}}, \bibinfo {author}
  {\bibfnamefont {R.~J.}\ \bibnamefont {Birgeneau}}, \ and\ \bibinfo {author}
  {\bibfnamefont {H.}~\bibnamefont {Zhang}},\ }\href {\doibase
  10.1103/PhysRevLett.98.247003} {\bibfield  {journal} {\bibinfo  {journal}
  {Phys. Rev. Lett.}\ }\textbf {\bibinfo {volume} {98}},\ \bibinfo {pages}
  {247003} (\bibinfo {year} {2007})}\BibitemShut {NoStop}%
\bibitem [{\citenamefont {Lipscombe}\ \emph {et~al.}(2007)\citenamefont
  {Lipscombe}, \citenamefont {Hayden}, \citenamefont {Vignolle}, \citenamefont
  {McMorrow},\ and\ \citenamefont {Perring}}]{LipscombePRL2007}%
  \BibitemOpen
  \bibfield  {author} {\bibinfo {author} {\bibfnamefont {O.~J.}\ \bibnamefont
  {Lipscombe}}, \bibinfo {author} {\bibfnamefont {S.~M.}\ \bibnamefont
  {Hayden}}, \bibinfo {author} {\bibfnamefont {B.}~\bibnamefont {Vignolle}},
  \bibinfo {author} {\bibfnamefont {D.~F.}\ \bibnamefont {McMorrow}}, \ and\
  \bibinfo {author} {\bibfnamefont {T.~G.}\ \bibnamefont {Perring}},\ }\href
  {\doibase 10.1103/PhysRevLett.99.067002} {\bibfield  {journal} {\bibinfo
  {journal} {Phys. Rev. Lett.}\ }\textbf {\bibinfo {volume} {99}},\ \bibinfo
  {pages} {067002} (\bibinfo {year} {2007})}\BibitemShut {NoStop}%
\bibitem [{\citenamefont {Stock}\ \emph {et~al.}(2010)\citenamefont {Stock},
  \citenamefont {Cowley}, \citenamefont {Buyers}, \citenamefont {Frost},
  \citenamefont {Taylor}, \citenamefont {Peets}, \citenamefont {Liang},
  \citenamefont {Bonn},\ and\ \citenamefont {Hardy}}]{StockPRB2010}%
  \BibitemOpen
  \bibfield  {author} {\bibinfo {author} {\bibfnamefont {C.}~\bibnamefont
  {Stock}}, \bibinfo {author} {\bibfnamefont {R.~A.}\ \bibnamefont {Cowley}},
  \bibinfo {author} {\bibfnamefont {W.~J.~L.}\ \bibnamefont {Buyers}}, \bibinfo
  {author} {\bibfnamefont {C.~D.}\ \bibnamefont {Frost}}, \bibinfo {author}
  {\bibfnamefont {J.~W.}\ \bibnamefont {Taylor}}, \bibinfo {author}
  {\bibfnamefont {D.}~\bibnamefont {Peets}}, \bibinfo {author} {\bibfnamefont
  {R.}~\bibnamefont {Liang}}, \bibinfo {author} {\bibfnamefont
  {D.}~\bibnamefont {Bonn}}, \ and\ \bibinfo {author} {\bibfnamefont {W.~N.}\
  \bibnamefont {Hardy}},\ }\href {\doibase 10.1103/PhysRevB.82.174505}
  {\bibfield  {journal} {\bibinfo  {journal} {Phys. Rev. B}\ }\textbf {\bibinfo
  {volume} {82}},\ \bibinfo {pages} {174505} (\bibinfo {year}
  {2010})}\BibitemShut {NoStop}%
\bibitem [{\citenamefont {Tranquada}\ \emph {et~al.}()\citenamefont
  {Tranquada}, \citenamefont {Xu},\ and\ \citenamefont
  {Zaliznyak}}]{TranquadaCondmat2013}%
  \BibitemOpen
  \bibfield  {author} {\bibinfo {author} {\bibfnamefont {J.~M.}\ \bibnamefont
  {Tranquada}}, \bibinfo {author} {\bibfnamefont {G.}~\bibnamefont {Xu}}, \
  and\ \bibinfo {author} {\bibfnamefont {I.~A.}\ \bibnamefont {Zaliznyak}},\
  }\href {http://lanl.arxiv.org/abs/1301.5888v1} {\enquote {\bibinfo {title}
  {Superconductivity, antiferromagnetism, and neutron scattering},}\ }\bibinfo
  {note} {ArXiv:1301.5888}\BibitemShut {NoStop}%
\bibitem [{\citenamefont {Lu}\ \emph {et~al.}(2005)\citenamefont {Lu},
  \citenamefont {Chabot-Couture}, \citenamefont {Zhao}, \citenamefont
  {Hancock}, \citenamefont {Kaneko}, \citenamefont {Vajk}, \citenamefont {Yu},
  \citenamefont {Grenier}, \citenamefont {Kim}, \citenamefont {Casa},
  \citenamefont {Gog},\ and\ \citenamefont {Greven}}]{LuPRL2005}%
  \BibitemOpen
  \bibfield  {author} {\bibinfo {author} {\bibfnamefont {L.}~\bibnamefont
  {Lu}}, \bibinfo {author} {\bibfnamefont {G.}~\bibnamefont {Chabot-Couture}},
  \bibinfo {author} {\bibfnamefont {X.}~\bibnamefont {Zhao}}, \bibinfo {author}
  {\bibfnamefont {J.~N.}\ \bibnamefont {Hancock}}, \bibinfo {author}
  {\bibfnamefont {N.}~\bibnamefont {Kaneko}}, \bibinfo {author} {\bibfnamefont
  {O.~P.}\ \bibnamefont {Vajk}}, \bibinfo {author} {\bibfnamefont
  {G.}~\bibnamefont {Yu}}, \bibinfo {author} {\bibfnamefont {S.}~\bibnamefont
  {Grenier}}, \bibinfo {author} {\bibfnamefont {Y.~J.}\ \bibnamefont {Kim}},
  \bibinfo {author} {\bibfnamefont {D.}~\bibnamefont {Casa}}, \bibinfo {author}
  {\bibfnamefont {T.}~\bibnamefont {Gog}}, \ and\ \bibinfo {author}
  {\bibfnamefont {M.}~\bibnamefont {Greven}},\ }\href {\doibase
  10.1103/PhysRevLett.95.217003} {\bibfield  {journal} {\bibinfo  {journal}
  {Phys. Rev. Lett.}\ }\textbf {\bibinfo {volume} {95}},\ \bibinfo {pages}
  {217003} (\bibinfo {year} {2005})}\BibitemShut {NoStop}%
\bibitem [{\citenamefont {Lu}\ \emph {et~al.}(2006)\citenamefont {Lu},
  \citenamefont {Hancock}, \citenamefont {Chabot-Couture}, \citenamefont
  {Ishii}, \citenamefont {Vajk}, \citenamefont {Yu}, \citenamefont {Mizuki},
  \citenamefont {Casa}, \citenamefont {Gog},\ and\ \citenamefont
  {Greven}}]{LuPRB2006}%
  \BibitemOpen
  \bibfield  {author} {\bibinfo {author} {\bibfnamefont {L.}~\bibnamefont
  {Lu}}, \bibinfo {author} {\bibfnamefont {J.~N.}\ \bibnamefont {Hancock}},
  \bibinfo {author} {\bibfnamefont {G.}~\bibnamefont {Chabot-Couture}},
  \bibinfo {author} {\bibfnamefont {K.}~\bibnamefont {Ishii}}, \bibinfo
  {author} {\bibfnamefont {O.~P.}\ \bibnamefont {Vajk}}, \bibinfo {author}
  {\bibfnamefont {G.}~\bibnamefont {Yu}}, \bibinfo {author} {\bibfnamefont
  {J.}~\bibnamefont {Mizuki}}, \bibinfo {author} {\bibfnamefont
  {D.}~\bibnamefont {Casa}}, \bibinfo {author} {\bibfnamefont {T.}~\bibnamefont
  {Gog}}, \ and\ \bibinfo {author} {\bibfnamefont {M.}~\bibnamefont {Greven}},\
  }\href {\doibase 10.1103/PhysRevB.74.224509} {\bibfield  {journal} {\bibinfo
  {journal} {Phys. Rev. B}\ }\textbf {\bibinfo {volume} {74}},\ \bibinfo
  {pages} {224509} (\bibinfo {year} {2006})}\BibitemShut {NoStop}%
\bibitem [{\citenamefont {Ament}\ \emph {et~al.}(2009)\citenamefont {Ament},
  \citenamefont {Ghiringhelli}, \citenamefont {Sala}, \citenamefont
  {Braicovich},\ and\ \citenamefont {van~den Brink}}]{AmentPRL2009}%
  \BibitemOpen
  \bibfield  {author} {\bibinfo {author} {\bibfnamefont {L.~J.~P.}\
  \bibnamefont {Ament}}, \bibinfo {author} {\bibfnamefont {G.}~\bibnamefont
  {Ghiringhelli}}, \bibinfo {author} {\bibfnamefont {M.~M.}\ \bibnamefont
  {Sala}}, \bibinfo {author} {\bibfnamefont {L.}~\bibnamefont {Braicovich}}, \
  and\ \bibinfo {author} {\bibfnamefont {J.}~\bibnamefont {van~den Brink}},\
  }\href {\doibase 10.1103/PhysRevLett.103.117003} {\bibfield  {journal}
  {\bibinfo  {journal} {Phys. Rev. Lett.}\ }\textbf {\bibinfo {volume} {103}},\
  \bibinfo {pages} {117003} (\bibinfo {year} {2009})}\BibitemShut {NoStop}%
\bibitem [{\citenamefont {Haverkort}(2010)}]{HaverkortPRL2010}%
  \BibitemOpen
  \bibfield  {author} {\bibinfo {author} {\bibfnamefont {M.~W.}\ \bibnamefont
  {Haverkort}},\ }\href {\doibase 10.1103/PhysRevLett.105.167404} {\bibfield
  {journal} {\bibinfo  {journal} {Phys. Rev. Lett.}\ }\textbf {\bibinfo
  {volume} {105}},\ \bibinfo {pages} {167404} (\bibinfo {year}
  {2010})}\BibitemShut {NoStop}%
\bibitem [{\citenamefont {Shastry}\ and\ \citenamefont
  {Shraiman}(1991)}]{ShastryIJMPB1991}%
  \BibitemOpen
  \bibfield  {author} {\bibinfo {author} {\bibfnamefont {B.~S.}\ \bibnamefont
  {Shastry}}\ and\ \bibinfo {author} {\bibfnamefont {B.~I.}\ \bibnamefont
  {Shraiman}},\ }\href {\doibase 10.1142/S0217979291000237} {\bibfield
  {journal} {\bibinfo  {journal} {Int. J. Modern Phy. B}\ }\textbf {\bibinfo
  {volume} {05}},\ \bibinfo {pages} {365} (\bibinfo {year} {1991})}\BibitemShut
  {NoStop}%
\bibitem [{\citenamefont {Devereaux}\ and\ \citenamefont
  {Hackl}(2007)}]{DevereauxRMP2007}%
  \BibitemOpen
  \bibfield  {author} {\bibinfo {author} {\bibfnamefont {T.~P.}\ \bibnamefont
  {Devereaux}}\ and\ \bibinfo {author} {\bibfnamefont {R.}~\bibnamefont
  {Hackl}},\ }\href {\doibase 10.1103/RevModPhys.79.175} {\bibfield  {journal}
  {\bibinfo  {journal} {Rev. Mod. Phys.}\ }\textbf {\bibinfo {volume} {79}},\
  \bibinfo {pages} {175} (\bibinfo {year} {2007})}\BibitemShut {NoStop}%
\bibitem [{\citenamefont {Yoshida}\ \emph {et~al.}(1992)\citenamefont
  {Yoshida}, \citenamefont {Tajima}, \citenamefont {Koshizuka}, \citenamefont
  {Tanaka}, \citenamefont {Uchida},\ and\ \citenamefont
  {Itoh}}]{YoshidaPRB1992}%
  \BibitemOpen
  \bibfield  {author} {\bibinfo {author} {\bibfnamefont {M.}~\bibnamefont
  {Yoshida}}, \bibinfo {author} {\bibfnamefont {S.}~\bibnamefont {Tajima}},
  \bibinfo {author} {\bibfnamefont {N.}~\bibnamefont {Koshizuka}}, \bibinfo
  {author} {\bibfnamefont {S.}~\bibnamefont {Tanaka}}, \bibinfo {author}
  {\bibfnamefont {S.}~\bibnamefont {Uchida}}, \ and\ \bibinfo {author}
  {\bibfnamefont {T.}~\bibnamefont {Itoh}},\ }\href {\doibase
  10.1103/PhysRevB.46.6505} {\bibfield  {journal} {\bibinfo  {journal} {Phys.
  Rev. B}\ }\textbf {\bibinfo {volume} {46}},\ \bibinfo {pages} {6505}
  (\bibinfo {year} {1992})}\BibitemShut {NoStop}%
\bibitem [{\citenamefont {Zhao}\ \emph {et~al.}(2006)\citenamefont {Zhao},
  \citenamefont {Yu}, \citenamefont {Cho}, \citenamefont {Chabot-Couture},
  \citenamefont {Bari\v{s}i\'{c}}, \citenamefont {Bourges}, \citenamefont
  {Kaneko}, \citenamefont {Li}, \citenamefont {Lu}, \citenamefont {Motoyama},
  \citenamefont {Vajk},\ and\ \citenamefont {Greven}}]{ZhaoAdvMater2006}%
  \BibitemOpen
  \bibfield  {author} {\bibinfo {author} {\bibfnamefont {X.}~\bibnamefont
  {Zhao}}, \bibinfo {author} {\bibfnamefont {G.}~\bibnamefont {Yu}}, \bibinfo
  {author} {\bibfnamefont {Y.-C.}\ \bibnamefont {Cho}}, \bibinfo {author}
  {\bibfnamefont {G.}~\bibnamefont {Chabot-Couture}}, \bibinfo {author}
  {\bibfnamefont {N.}~\bibnamefont {Bari\v{s}i\'{c}}}, \bibinfo {author}
  {\bibfnamefont {P.}~\bibnamefont {Bourges}}, \bibinfo {author} {\bibfnamefont
  {N.}~\bibnamefont {Kaneko}}, \bibinfo {author} {\bibfnamefont
  {Y.}~\bibnamefont {Li}}, \bibinfo {author} {\bibfnamefont {L.}~\bibnamefont
  {Lu}}, \bibinfo {author} {\bibfnamefont {E.~M.}\ \bibnamefont {Motoyama}},
  \bibinfo {author} {\bibfnamefont {O.~P.}\ \bibnamefont {Vajk}}, \ and\
  \bibinfo {author} {\bibfnamefont {M.}~\bibnamefont {Greven}},\ }\href@noop {}
  {\bibfield  {journal} {\bibinfo  {journal} {Adv. Mater.}\ }\textbf {\bibinfo
  {volume} {18}},\ \bibinfo {pages} {3243} (\bibinfo {year}
  {2006})}\BibitemShut {NoStop}%
\bibitem [{\citenamefont {Bari\v{s}i\'{c}}\ \emph {et~al.}(2008)\citenamefont
  {Bari\v{s}i\'{c}}, \citenamefont {Li}, \citenamefont {Zhao}, \citenamefont
  {Cho}, \citenamefont {Chabot-Couture}, \citenamefont {Yu},\ and\
  \citenamefont {Greven}}]{BarisicPRB2008}%
  \BibitemOpen
  \bibfield  {author} {\bibinfo {author} {\bibfnamefont {N.}~\bibnamefont
  {Bari\v{s}i\'{c}}}, \bibinfo {author} {\bibfnamefont {Y.}~\bibnamefont {Li}},
  \bibinfo {author} {\bibfnamefont {X.}~\bibnamefont {Zhao}}, \bibinfo {author}
  {\bibfnamefont {Y.-C.}\ \bibnamefont {Cho}}, \bibinfo {author} {\bibfnamefont
  {G.}~\bibnamefont {Chabot-Couture}}, \bibinfo {author} {\bibfnamefont
  {G.}~\bibnamefont {Yu}}, \ and\ \bibinfo {author} {\bibfnamefont
  {M.}~\bibnamefont {Greven}},\ }\href@noop {} {\bibfield  {journal} {\bibinfo
  {journal} {Phys. Rev. B}\ }\textbf {\bibinfo {volume} {78}},\ \bibinfo {eid}
  {054518} (\bibinfo {year} {2008})}\BibitemShut {NoStop}%
\bibitem [{\citenamefont {Tallon}\ \emph {et~al.}(1995)\citenamefont {Tallon},
  \citenamefont {Bernhard}, \citenamefont {Shaked}, \citenamefont {Hitterman},\
  and\ \citenamefont {Jorgensen}}]{TallonPRB1995}%
  \BibitemOpen
  \bibfield  {author} {\bibinfo {author} {\bibfnamefont {J.~L.}\ \bibnamefont
  {Tallon}}, \bibinfo {author} {\bibfnamefont {C.}~\bibnamefont {Bernhard}},
  \bibinfo {author} {\bibfnamefont {H.}~\bibnamefont {Shaked}}, \bibinfo
  {author} {\bibfnamefont {R.~L.}\ \bibnamefont {Hitterman}}, \ and\ \bibinfo
  {author} {\bibfnamefont {J.~D.}\ \bibnamefont {Jorgensen}},\ }\href {\doibase
  10.1103/PhysRevB.51.12911} {\bibfield  {journal} {\bibinfo  {journal} {Phys.
  Rev. B}\ }\textbf {\bibinfo {volume} {51}},\ \bibinfo {pages} {12911}
  (\bibinfo {year} {1995})}\BibitemShut {NoStop}%
\bibitem [{\citenamefont {Le~Tacon}\ \emph {et~al.}(2006)\citenamefont
  {Le~Tacon}, \citenamefont {Sacuto}, \citenamefont {Georges}, \citenamefont
  {Kotliar}, \citenamefont {Gallais}, \citenamefont {Colson},\ and\
  \citenamefont {Forget}}]{LeTaconNatPhys2006}%
  \BibitemOpen
  \bibfield  {author} {\bibinfo {author} {\bibfnamefont {M.}~\bibnamefont
  {Le~Tacon}}, \bibinfo {author} {\bibfnamefont {A.}~\bibnamefont {Sacuto}},
  \bibinfo {author} {\bibfnamefont {A.}~\bibnamefont {Georges}}, \bibinfo
  {author} {\bibfnamefont {G.}~\bibnamefont {Kotliar}}, \bibinfo {author}
  {\bibfnamefont {Y.}~\bibnamefont {Gallais}}, \bibinfo {author} {\bibfnamefont
  {D.}~\bibnamefont {Colson}}, \ and\ \bibinfo {author} {\bibfnamefont
  {A.}~\bibnamefont {Forget}},\ }\href@noop {} {\bibfield  {journal} {\bibinfo
  {journal} {Nature Phys.}\ }\textbf {\bibinfo {volume} {2}},\ \bibinfo {pages}
  {537} (\bibinfo {year} {2006})}\BibitemShut {NoStop}%
\bibitem [{\citenamefont {Blanc}\ \emph {et~al.}(2010)\citenamefont {Blanc},
  \citenamefont {Gallais}, \citenamefont {Cazayous}, \citenamefont {M\'easson},
  \citenamefont {Sacuto}, \citenamefont {Georges}, \citenamefont {Wen},
  \citenamefont {Xu}, \citenamefont {Gu},\ and\ \citenamefont
  {Colson}}]{BlancPRB2010}%
  \BibitemOpen
  \bibfield  {author} {\bibinfo {author} {\bibfnamefont {S.}~\bibnamefont
  {Blanc}}, \bibinfo {author} {\bibfnamefont {Y.}~\bibnamefont {Gallais}},
  \bibinfo {author} {\bibfnamefont {M.}~\bibnamefont {Cazayous}}, \bibinfo
  {author} {\bibfnamefont {M.~A.}\ \bibnamefont {M\'easson}}, \bibinfo {author}
  {\bibfnamefont {A.}~\bibnamefont {Sacuto}}, \bibinfo {author} {\bibfnamefont
  {A.}~\bibnamefont {Georges}}, \bibinfo {author} {\bibfnamefont {J.~S.}\
  \bibnamefont {Wen}}, \bibinfo {author} {\bibfnamefont {Z.~J.}\ \bibnamefont
  {Xu}}, \bibinfo {author} {\bibfnamefont {G.~D.}\ \bibnamefont {Gu}}, \ and\
  \bibinfo {author} {\bibfnamefont {D.}~\bibnamefont {Colson}},\ }\href
  {\doibase 10.1103/PhysRevB.82.144516} {\bibfield  {journal} {\bibinfo
  {journal} {Phys. Rev. B}\ }\textbf {\bibinfo {volume} {82}},\ \bibinfo
  {pages} {144516} (\bibinfo {year} {2010})}\BibitemShut {NoStop}%
\bibitem [{\citenamefont {Munnikes}\ \emph {et~al.}(2011)\citenamefont
  {Munnikes}, \citenamefont {Muschler}, \citenamefont {Venturini},
  \citenamefont {Tassini}, \citenamefont {Prestel}, \citenamefont {Ono},
  \citenamefont {Ando}, \citenamefont {Peets}, \citenamefont {Hardy},
  \citenamefont {Liang}, \citenamefont {Bonn}, \citenamefont {Damascelli},
  \citenamefont {Eisaki}, \citenamefont {Greven}, \citenamefont {Erb},\ and\
  \citenamefont {Hackl}}]{MunnikesPRB2011}%
  \BibitemOpen
  \bibfield  {author} {\bibinfo {author} {\bibfnamefont {N.}~\bibnamefont
  {Munnikes}}, \bibinfo {author} {\bibfnamefont {B.}~\bibnamefont {Muschler}},
  \bibinfo {author} {\bibfnamefont {F.}~\bibnamefont {Venturini}}, \bibinfo
  {author} {\bibfnamefont {L.}~\bibnamefont {Tassini}}, \bibinfo {author}
  {\bibfnamefont {W.}~\bibnamefont {Prestel}}, \bibinfo {author} {\bibfnamefont
  {S.}~\bibnamefont {Ono}}, \bibinfo {author} {\bibfnamefont {Y.}~\bibnamefont
  {Ando}}, \bibinfo {author} {\bibfnamefont {D.~C.}\ \bibnamefont {Peets}},
  \bibinfo {author} {\bibfnamefont {W.~N.}\ \bibnamefont {Hardy}}, \bibinfo
  {author} {\bibfnamefont {R.}~\bibnamefont {Liang}}, \bibinfo {author}
  {\bibfnamefont {D.~A.}\ \bibnamefont {Bonn}}, \bibinfo {author}
  {\bibfnamefont {A.}~\bibnamefont {Damascelli}}, \bibinfo {author}
  {\bibfnamefont {H.}~\bibnamefont {Eisaki}}, \bibinfo {author} {\bibfnamefont
  {M.}~\bibnamefont {Greven}}, \bibinfo {author} {\bibfnamefont
  {A.}~\bibnamefont {Erb}}, \ and\ \bibinfo {author} {\bibfnamefont
  {R.}~\bibnamefont {Hackl}},\ }\href {\doibase 10.1103/PhysRevB.84.144523}
  {\bibfield  {journal} {\bibinfo  {journal} {Phys. Rev. B}\ }\textbf {\bibinfo
  {volume} {84}},\ \bibinfo {pages} {144523} (\bibinfo {year}
  {2011})}\BibitemShut {NoStop}%
\bibitem [{\citenamefont {Gallais}\ \emph {et~al.}(2005)\citenamefont
  {Gallais}, \citenamefont {Sacuto}, \citenamefont {Devereaux},\ and\
  \citenamefont {Colson}}]{GallaisPRB2005}%
  \BibitemOpen
  \bibfield  {author} {\bibinfo {author} {\bibfnamefont {Y.}~\bibnamefont
  {Gallais}}, \bibinfo {author} {\bibfnamefont {A.}~\bibnamefont {Sacuto}},
  \bibinfo {author} {\bibfnamefont {T.~P.}\ \bibnamefont {Devereaux}}, \ and\
  \bibinfo {author} {\bibfnamefont {D.}~\bibnamefont {Colson}},\ }\href
  {\doibase 10.1103/PhysRevB.71.012506} {\bibfield  {journal} {\bibinfo
  {journal} {Phys. Rev. B}\ }\textbf {\bibinfo {volume} {71}},\ \bibinfo
  {pages} {012506} (\bibinfo {year} {2005})}\BibitemShut {NoStop}%
\bibitem [{\citenamefont {Sakai}\ \emph {et~al.}()\citenamefont {Sakai},
  \citenamefont {Blanc}, \citenamefont {Civelli}, \citenamefont {Gallais},
  \citenamefont {Cazayous}, \citenamefont {M\'easson}, \citenamefont {Wen},
  \citenamefont {Xu}, \citenamefont {Gu}, \citenamefont {Sangiovanni},
  \citenamefont {Motome}, \citenamefont {Held}, \citenamefont {Sacuto},
  \citenamefont {Georges},\ and\ \citenamefont {Imada}}]{SakaiCondmat2012}%
  \BibitemOpen
  \bibfield  {author} {\bibinfo {author} {\bibfnamefont {S.}~\bibnamefont
  {Sakai}}, \bibinfo {author} {\bibfnamefont {S.}~\bibnamefont {Blanc}},
  \bibinfo {author} {\bibfnamefont {M.}~\bibnamefont {Civelli}}, \bibinfo
  {author} {\bibfnamefont {Y.}~\bibnamefont {Gallais}}, \bibinfo {author}
  {\bibfnamefont {M.}~\bibnamefont {Cazayous}}, \bibinfo {author}
  {\bibfnamefont {M.-A.}\ \bibnamefont {M\'easson}}, \bibinfo {author}
  {\bibfnamefont {J.~S.}\ \bibnamefont {Wen}}, \bibinfo {author} {\bibfnamefont
  {Z.~J.}\ \bibnamefont {Xu}}, \bibinfo {author} {\bibfnamefont {G.~D.}\
  \bibnamefont {Gu}}, \bibinfo {author} {\bibfnamefont {G.}~\bibnamefont
  {Sangiovanni}}, \bibinfo {author} {\bibfnamefont {Y.}~\bibnamefont {Motome}},
  \bibinfo {author} {\bibfnamefont {K.}~\bibnamefont {Held}}, \bibinfo {author}
  {\bibfnamefont {A.}~\bibnamefont {Sacuto}}, \bibinfo {author} {\bibfnamefont
  {A.}~\bibnamefont {Georges}}, \ and\ \bibinfo {author} {\bibfnamefont
  {M.}~\bibnamefont {Imada}},\ }\href@noop {} {\enquote {\bibinfo {title}
  {Exploring the dark sie of cuprate superconductors: $s$-wave symmetry of the
  pseudogap},}\ }\bibinfo {note} {ArXiv:1207.5070}\BibitemShut {NoStop}%
\bibitem [{\citenamefont {Barlas}\ and\ \citenamefont
  {Varma}()}]{BarlasCondmat2013}%
  \BibitemOpen
  \bibfield  {author} {\bibinfo {author} {\bibfnamefont {Y.}~\bibnamefont
  {Barlas}}\ and\ \bibinfo {author} {\bibfnamefont {C.~M.}\ \bibnamefont
  {Varma}},\ }\href@noop {} {\enquote {\bibinfo {title} {Higgs bosons in
  $d$-wave superconductors},}\ }\bibinfo {note} {ArXiv:1206.0400}\BibitemShut
  {NoStop}%
\bibitem [{\citenamefont {Tanaka}\ \emph {et~al.}(2006)\citenamefont {Tanaka},
  \citenamefont {Lee}, \citenamefont {Lu}, \citenamefont {Fujimori},
  \citenamefont {Fujii}, \citenamefont {Risdiana}, \citenamefont {Terasaki},
  \citenamefont {Scalapino}, \citenamefont {Devereaux}, \citenamefont
  {Hussain},\ and\ \citenamefont {Shen}}]{TanakaScience2006}%
  \BibitemOpen
  \bibfield  {author} {\bibinfo {author} {\bibfnamefont {K.}~\bibnamefont
  {Tanaka}}, \bibinfo {author} {\bibfnamefont {W.~S.}\ \bibnamefont {Lee}},
  \bibinfo {author} {\bibfnamefont {D.~H.}\ \bibnamefont {Lu}}, \bibinfo
  {author} {\bibfnamefont {A.}~\bibnamefont {Fujimori}}, \bibinfo {author}
  {\bibfnamefont {T.}~\bibnamefont {Fujii}}, \bibinfo {author} {\bibnamefont
  {Risdiana}}, \bibinfo {author} {\bibfnamefont {I.}~\bibnamefont {Terasaki}},
  \bibinfo {author} {\bibfnamefont {D.~J.}\ \bibnamefont {Scalapino}}, \bibinfo
  {author} {\bibfnamefont {T.~P.}\ \bibnamefont {Devereaux}}, \bibinfo {author}
  {\bibfnamefont {Z.}~\bibnamefont {Hussain}}, \ and\ \bibinfo {author}
  {\bibfnamefont {Z.-X.}\ \bibnamefont {Shen}},\ }\href {\doibase
  10.1126/science.1133411} {\bibfield  {journal} {\bibinfo  {journal}
  {Science}\ }\textbf {\bibinfo {volume} {314}},\ \bibinfo {pages} {1910}
  (\bibinfo {year} {2006})}\BibitemShut {NoStop}%
\bibitem [{\citenamefont {Vishik}\ \emph {et~al.}(2012)\citenamefont {Vishik},
  \citenamefont {Hashimoto}, \citenamefont {He}, \citenamefont {Lee},
  \citenamefont {Schmitt}, \citenamefont {Lu}, \citenamefont {Moore},
  \citenamefont {Zhang}, \citenamefont {Meevasana}, \citenamefont {Sasagawa},
  \citenamefont {Uchida}, \citenamefont {Fujita}, \citenamefont {Ishida},
  \citenamefont {Ishikado}, \citenamefont {Yoshida}, \citenamefont {Eisaki},
  \citenamefont {Hussain}, \citenamefont {Devereaux},\ and\ \citenamefont
  {Shen}}]{VishikPNAS2012}%
  \BibitemOpen
  \bibfield  {author} {\bibinfo {author} {\bibfnamefont {I.~M.}\ \bibnamefont
  {Vishik}}, \bibinfo {author} {\bibfnamefont {M.}~\bibnamefont {Hashimoto}},
  \bibinfo {author} {\bibfnamefont {R.-H.}\ \bibnamefont {He}}, \bibinfo
  {author} {\bibfnamefont {W.-S.}\ \bibnamefont {Lee}}, \bibinfo {author}
  {\bibfnamefont {F.}~\bibnamefont {Schmitt}}, \bibinfo {author} {\bibfnamefont
  {D.}~\bibnamefont {Lu}}, \bibinfo {author} {\bibfnamefont {R.~G.}\
  \bibnamefont {Moore}}, \bibinfo {author} {\bibfnamefont {C.}~\bibnamefont
  {Zhang}}, \bibinfo {author} {\bibfnamefont {W.}~\bibnamefont {Meevasana}},
  \bibinfo {author} {\bibfnamefont {T.}~\bibnamefont {Sasagawa}}, \bibinfo
  {author} {\bibfnamefont {S.}~\bibnamefont {Uchida}}, \bibinfo {author}
  {\bibfnamefont {K.}~\bibnamefont {Fujita}}, \bibinfo {author} {\bibfnamefont
  {S.}~\bibnamefont {Ishida}}, \bibinfo {author} {\bibfnamefont
  {M.}~\bibnamefont {Ishikado}}, \bibinfo {author} {\bibfnamefont
  {Y.}~\bibnamefont {Yoshida}}, \bibinfo {author} {\bibfnamefont
  {H.}~\bibnamefont {Eisaki}}, \bibinfo {author} {\bibfnamefont
  {Z.}~\bibnamefont {Hussain}}, \bibinfo {author} {\bibfnamefont {T.~P.}\
  \bibnamefont {Devereaux}}, \ and\ \bibinfo {author} {\bibfnamefont {Z.-X.}\
  \bibnamefont {Shen}},\ }\href {\doibase 10.1073/pnas.1209471109} {\bibfield
  {journal} {\bibinfo  {journal} {Proc. Nat. Acad. Sci.}\ }\textbf {\bibinfo
  {volume} {109}},\ \bibinfo {pages} {18332} (\bibinfo {year}
  {2012})}\BibitemShut {NoStop}%
\bibitem [{\citenamefont {Chubukov}\ and\ \citenamefont
  {Frenkel}(1995{\natexlab{a}})}]{ChubukovPRL1995}%
  \BibitemOpen
  \bibfield  {author} {\bibinfo {author} {\bibfnamefont {A.~V.}\ \bibnamefont
  {Chubukov}}\ and\ \bibinfo {author} {\bibfnamefont {D.~M.}\ \bibnamefont
  {Frenkel}},\ }\href {\doibase 10.1103/PhysRevLett.74.3057} {\bibfield
  {journal} {\bibinfo  {journal} {Phys. Rev. Lett.}\ }\textbf {\bibinfo
  {volume} {74}},\ \bibinfo {pages} {3057} (\bibinfo {year}
  {1995}{\natexlab{a}})}\BibitemShut {NoStop}%
\bibitem [{\citenamefont {Chubukov}\ and\ \citenamefont
  {Frenkel}(1995{\natexlab{b}})}]{ChubukovPRB1995}%
  \BibitemOpen
  \bibfield  {author} {\bibinfo {author} {\bibfnamefont {A.~V.}\ \bibnamefont
  {Chubukov}}\ and\ \bibinfo {author} {\bibfnamefont {D.~M.}\ \bibnamefont
  {Frenkel}},\ }\href {\doibase 10.1103/PhysRevB.52.9760} {\bibfield  {journal}
  {\bibinfo  {journal} {Phys. Rev. B}\ }\textbf {\bibinfo {volume} {52}},\
  \bibinfo {pages} {9760} (\bibinfo {year} {1995}{\natexlab{b}})}\BibitemShut
  {NoStop}%
\bibitem [{\citenamefont {Hanamura}\ \emph {et~al.}(2000)\citenamefont
  {Hanamura}, \citenamefont {Dan},\ and\ \citenamefont
  {Tanabe}}]{HanamuraPRB2000}%
  \BibitemOpen
  \bibfield  {author} {\bibinfo {author} {\bibfnamefont {E.}~\bibnamefont
  {Hanamura}}, \bibinfo {author} {\bibfnamefont {N.~T.}\ \bibnamefont {Dan}}, \
  and\ \bibinfo {author} {\bibfnamefont {Y.}~\bibnamefont {Tanabe}},\ }\href
  {\doibase 10.1103/PhysRevB.62.7033} {\bibfield  {journal} {\bibinfo
  {journal} {Phys. Rev. B}\ }\textbf {\bibinfo {volume} {62}},\ \bibinfo
  {pages} {7033} (\bibinfo {year} {2000})}\BibitemShut {NoStop}%
\bibitem [{\citenamefont {Tohyama}\ \emph {et~al.}(2002)\citenamefont
  {Tohyama}, \citenamefont {Onodera}, \citenamefont {Tsutsui},\ and\
  \citenamefont {Maekawa}}]{TohyamaPRL2002}%
  \BibitemOpen
  \bibfield  {author} {\bibinfo {author} {\bibfnamefont {T.}~\bibnamefont
  {Tohyama}}, \bibinfo {author} {\bibfnamefont {H.}~\bibnamefont {Onodera}},
  \bibinfo {author} {\bibfnamefont {K.}~\bibnamefont {Tsutsui}}, \ and\
  \bibinfo {author} {\bibfnamefont {S.}~\bibnamefont {Maekawa}},\ }\href
  {\doibase 10.1103/PhysRevLett.89.257405} {\bibfield  {journal} {\bibinfo
  {journal} {Phys. Rev. Lett.}\ }\textbf {\bibinfo {volume} {89}},\ \bibinfo
  {pages} {257405} (\bibinfo {year} {2002})}\BibitemShut {NoStop}%
\bibitem [{\citenamefont {Kup\v{c}i\'{c}}(2011)}]{KupcicJRS2011}%
  \BibitemOpen
  \bibfield  {author} {\bibinfo {author} {\bibfnamefont {I.}~\bibnamefont
  {Kup\v{c}i\'{c}}},\ }\href {\doibase 10.1002/jrs.2792} {\bibfield  {journal}
  {\bibinfo  {journal} {Journal of Raman Spectroscopy}\ }\textbf {\bibinfo
  {volume} {42}},\ \bibinfo {pages} {998} (\bibinfo {year} {2011})}\BibitemShut
  {NoStop}%
\bibitem [{\citenamefont {Uchida}\ \emph {et~al.}(1991)\citenamefont {Uchida},
  \citenamefont {Ido}, \citenamefont {Takagi}, \citenamefont {Arima},
  \citenamefont {Tokura},\ and\ \citenamefont {Tajima}}]{UchidaPRB1991}%
  \BibitemOpen
  \bibfield  {author} {\bibinfo {author} {\bibfnamefont {S.}~\bibnamefont
  {Uchida}}, \bibinfo {author} {\bibfnamefont {T.}~\bibnamefont {Ido}},
  \bibinfo {author} {\bibfnamefont {H.}~\bibnamefont {Takagi}}, \bibinfo
  {author} {\bibfnamefont {T.}~\bibnamefont {Arima}}, \bibinfo {author}
  {\bibfnamefont {Y.}~\bibnamefont {Tokura}}, \ and\ \bibinfo {author}
  {\bibfnamefont {S.}~\bibnamefont {Tajima}},\ }\href {\doibase
  10.1103/PhysRevB.43.7942} {\bibfield  {journal} {\bibinfo  {journal} {Phys.
  Rev. B}\ }\textbf {\bibinfo {volume} {43}},\ \bibinfo {pages} {7942}
  (\bibinfo {year} {1991})}\BibitemShut {NoStop}%
\bibitem [{\citenamefont {Giannetti}\ \emph {et~al.}(2011)\citenamefont
  {Giannetti}, \citenamefont {Cilento}, \citenamefont {Dal~Conte},
  \citenamefont {Coslovich}, \citenamefont {Ferrini}, \citenamefont
  {Molegraaf}, \citenamefont {Raichle}, \citenamefont {Liang}, \citenamefont
  {Eisaki}, \citenamefont {Greven}, \citenamefont {Damascelli}, \citenamefont
  {van~der Marel},\ and\ \citenamefont {Parmigiani}}]{GiannettiNatComm2011}%
  \BibitemOpen
  \bibfield  {author} {\bibinfo {author} {\bibfnamefont {C.}~\bibnamefont
  {Giannetti}}, \bibinfo {author} {\bibfnamefont {F.}~\bibnamefont {Cilento}},
  \bibinfo {author} {\bibfnamefont {S.}~\bibnamefont {Dal~Conte}}, \bibinfo
  {author} {\bibfnamefont {G.}~\bibnamefont {Coslovich}}, \bibinfo {author}
  {\bibfnamefont {G.}~\bibnamefont {Ferrini}}, \bibinfo {author} {\bibfnamefont
  {H.}~\bibnamefont {Molegraaf}}, \bibinfo {author} {\bibfnamefont
  {M.}~\bibnamefont {Raichle}}, \bibinfo {author} {\bibfnamefont
  {R.}~\bibnamefont {Liang}}, \bibinfo {author} {\bibfnamefont
  {H.}~\bibnamefont {Eisaki}}, \bibinfo {author} {\bibfnamefont
  {M.}~\bibnamefont {Greven}}, \bibinfo {author} {\bibfnamefont
  {A.}~\bibnamefont {Damascelli}}, \bibinfo {author} {\bibfnamefont
  {D.}~\bibnamefont {van~der Marel}}, \ and\ \bibinfo {author} {\bibfnamefont
  {F.}~\bibnamefont {Parmigiani}},\ }\href@noop {} {\bibfield  {journal}
  {\bibinfo  {journal} {Nature Communications}\ }\textbf {\bibinfo {volume}
  {2}},\ \bibinfo {pages} {353} (\bibinfo {year} {2011})}\BibitemShut {NoStop}%
\bibitem [{\citenamefont {Lorenzana}\ and\ \citenamefont
  {Yu}(1993)}]{LorenzanaPRL1993}%
  \BibitemOpen
  \bibfield  {author} {\bibinfo {author} {\bibfnamefont {J.}~\bibnamefont
  {Lorenzana}}\ and\ \bibinfo {author} {\bibfnamefont {L.}~\bibnamefont {Yu}},\
  }\href {\doibase 10.1103/PhysRevLett.70.861} {\bibfield  {journal} {\bibinfo
  {journal} {Phys. Rev. Lett.}\ }\textbf {\bibinfo {volume} {70}},\ \bibinfo
  {pages} {861} (\bibinfo {year} {1993})}\BibitemShut {NoStop}%
\bibitem [{\citenamefont {Schneider}\ \emph {et~al.}(2005)\citenamefont
  {Schneider}, \citenamefont {Unger}, \citenamefont {Mitdank}, \citenamefont
  {M\"uller}, \citenamefont {Krapf}, \citenamefont {Rogaschewski},
  \citenamefont {Dwelk}, \citenamefont {Janowitz},\ and\ \citenamefont
  {Manzke}}]{SchneiderPRB2005}%
  \BibitemOpen
  \bibfield  {author} {\bibinfo {author} {\bibfnamefont {M.}~\bibnamefont
  {Schneider}}, \bibinfo {author} {\bibfnamefont {R.-S.}\ \bibnamefont
  {Unger}}, \bibinfo {author} {\bibfnamefont {R.}~\bibnamefont {Mitdank}},
  \bibinfo {author} {\bibfnamefont {R.}~\bibnamefont {M\"uller}}, \bibinfo
  {author} {\bibfnamefont {A.}~\bibnamefont {Krapf}}, \bibinfo {author}
  {\bibfnamefont {S.}~\bibnamefont {Rogaschewski}}, \bibinfo {author}
  {\bibfnamefont {H.}~\bibnamefont {Dwelk}}, \bibinfo {author} {\bibfnamefont
  {C.}~\bibnamefont {Janowitz}}, \ and\ \bibinfo {author} {\bibfnamefont
  {R.}~\bibnamefont {Manzke}},\ }\href {\doibase 10.1103/PhysRevB.72.014504}
  {\bibfield  {journal} {\bibinfo  {journal} {Phys. Rev. B}\ }\textbf {\bibinfo
  {volume} {72}},\ \bibinfo {pages} {014504} (\bibinfo {year}
  {2005})}\BibitemShut {NoStop}%
\bibitem [{\citenamefont {Peets}\ \emph {et~al.}(2009)\citenamefont {Peets},
  \citenamefont {Hawthorn}, \citenamefont {Shen}, \citenamefont {Kim},
  \citenamefont {Ellis}, \citenamefont {Zhang}, \citenamefont {Komiya},
  \citenamefont {Ando}, \citenamefont {Sawatzky}, \citenamefont {Liang},
  \citenamefont {Bonn},\ and\ \citenamefont {Hardy}}]{PeetsPRL2009}%
  \BibitemOpen
  \bibfield  {author} {\bibinfo {author} {\bibfnamefont {D.~C.}\ \bibnamefont
  {Peets}}, \bibinfo {author} {\bibfnamefont {D.~G.}\ \bibnamefont {Hawthorn}},
  \bibinfo {author} {\bibfnamefont {K.~M.}\ \bibnamefont {Shen}}, \bibinfo
  {author} {\bibfnamefont {Y.-J.}\ \bibnamefont {Kim}}, \bibinfo {author}
  {\bibfnamefont {D.~S.}\ \bibnamefont {Ellis}}, \bibinfo {author}
  {\bibfnamefont {H.}~\bibnamefont {Zhang}}, \bibinfo {author} {\bibfnamefont
  {S.}~\bibnamefont {Komiya}}, \bibinfo {author} {\bibfnamefont
  {Y.}~\bibnamefont {Ando}}, \bibinfo {author} {\bibfnamefont {G.~A.}\
  \bibnamefont {Sawatzky}}, \bibinfo {author} {\bibfnamefont {R.}~\bibnamefont
  {Liang}}, \bibinfo {author} {\bibfnamefont {D.~A.}\ \bibnamefont {Bonn}}, \
  and\ \bibinfo {author} {\bibfnamefont {W.~N.}\ \bibnamefont {Hardy}},\ }\href
  {\doibase 10.1103/PhysRevLett.103.087402} {\bibfield  {journal} {\bibinfo
  {journal} {Phys. Rev. Lett.}\ }\textbf {\bibinfo {volume} {103}},\ \bibinfo
  {pages} {087402} (\bibinfo {year} {2009})}\BibitemShut {NoStop}%
\bibitem [{\citenamefont {Chen}\ \emph {et~al.}(1992)\citenamefont {Chen},
  \citenamefont {Tjeng}, \citenamefont {Kwo}, \citenamefont {Kao},
  \citenamefont {Rudolf}, \citenamefont {Sette},\ and\ \citenamefont
  {Fleming}}]{ChenPRL1992}%
  \BibitemOpen
  \bibfield  {author} {\bibinfo {author} {\bibfnamefont {C.~T.}\ \bibnamefont
  {Chen}}, \bibinfo {author} {\bibfnamefont {L.~H.}\ \bibnamefont {Tjeng}},
  \bibinfo {author} {\bibfnamefont {J.}~\bibnamefont {Kwo}}, \bibinfo {author}
  {\bibfnamefont {H.~L.}\ \bibnamefont {Kao}}, \bibinfo {author} {\bibfnamefont
  {P.}~\bibnamefont {Rudolf}}, \bibinfo {author} {\bibfnamefont
  {F.}~\bibnamefont {Sette}}, \ and\ \bibinfo {author} {\bibfnamefont {R.~M.}\
  \bibnamefont {Fleming}},\ }\href {\doibase 10.1103/PhysRevLett.68.2543}
  {\bibfield  {journal} {\bibinfo  {journal} {Phys. Rev. Lett.}\ }\textbf
  {\bibinfo {volume} {68}},\ \bibinfo {pages} {2543} (\bibinfo {year}
  {1992})}\BibitemShut {NoStop}%
\bibitem [{\citenamefont {Merz}\ \emph {et~al.}(1998)\citenamefont {Merz},
  \citenamefont {N\"ucker}, \citenamefont {Schweiss}, \citenamefont
  {Schuppler}, \citenamefont {Chen}, \citenamefont {Chakarian}, \citenamefont
  {Freeland}, \citenamefont {Idzerda}, \citenamefont {Kl\"aser}, \citenamefont
  {M\"uller-Vogt},\ and\ \citenamefont {Wolf}}]{MerzPRL1998}%
  \BibitemOpen
  \bibfield  {author} {\bibinfo {author} {\bibfnamefont {M.}~\bibnamefont
  {Merz}}, \bibinfo {author} {\bibfnamefont {N.}~\bibnamefont {N\"ucker}},
  \bibinfo {author} {\bibfnamefont {P.}~\bibnamefont {Schweiss}}, \bibinfo
  {author} {\bibfnamefont {S.}~\bibnamefont {Schuppler}}, \bibinfo {author}
  {\bibfnamefont {C.~T.}\ \bibnamefont {Chen}}, \bibinfo {author}
  {\bibfnamefont {V.}~\bibnamefont {Chakarian}}, \bibinfo {author}
  {\bibfnamefont {J.}~\bibnamefont {Freeland}}, \bibinfo {author}
  {\bibfnamefont {Y.~U.}\ \bibnamefont {Idzerda}}, \bibinfo {author}
  {\bibfnamefont {M.}~\bibnamefont {Kl\"aser}}, \bibinfo {author}
  {\bibfnamefont {G.}~\bibnamefont {M\"uller-Vogt}}, \ and\ \bibinfo {author}
  {\bibfnamefont {T.}~\bibnamefont {Wolf}},\ }\href {\doibase
  10.1103/PhysRevLett.80.5192} {\bibfield  {journal} {\bibinfo  {journal}
  {Phys. Rev. Lett.}\ }\textbf {\bibinfo {volume} {80}},\ \bibinfo {pages}
  {5192} (\bibinfo {year} {1998})}\BibitemShut {NoStop}%
\bibitem [{\citenamefont {Sedrakyan}\ and\ \citenamefont
  {Chubukov}(2010)}]{SedrakyanPRB2010}%
  \BibitemOpen
  \bibfield  {author} {\bibinfo {author} {\bibfnamefont {T.~A.}\ \bibnamefont
  {Sedrakyan}}\ and\ \bibinfo {author} {\bibfnamefont {A.~V.}\ \bibnamefont
  {Chubukov}},\ }\href {\doibase 10.1103/PhysRevB.81.174536} {\bibfield
  {journal} {\bibinfo  {journal} {Phys. Rev. B}\ }\textbf {\bibinfo {volume}
  {81}},\ \bibinfo {pages} {174536} (\bibinfo {year} {2010})}\BibitemShut
  {NoStop}%
\bibitem [{\citenamefont {Sakakibara}\ \emph {et~al.}(2012)\citenamefont
  {Sakakibara}, \citenamefont {Usui}, \citenamefont {Kuroki}, \citenamefont
  {Arita},\ and\ \citenamefont {Aoki}}]{SakakibaraPRB2012}%
  \BibitemOpen
  \bibfield  {author} {\bibinfo {author} {\bibfnamefont {H.}~\bibnamefont
  {Sakakibara}}, \bibinfo {author} {\bibfnamefont {H.}~\bibnamefont {Usui}},
  \bibinfo {author} {\bibfnamefont {K.}~\bibnamefont {Kuroki}}, \bibinfo
  {author} {\bibfnamefont {R.}~\bibnamefont {Arita}}, \ and\ \bibinfo {author}
  {\bibfnamefont {H.}~\bibnamefont {Aoki}},\ }\href {\doibase
  10.1103/PhysRevB.85.064501} {\bibfield  {journal} {\bibinfo  {journal} {Phys.
  Rev. B}\ }\textbf {\bibinfo {volume} {85}},\ \bibinfo {pages} {064501}
  (\bibinfo {year} {2012})}\BibitemShut {NoStop}%
\bibitem [{\citenamefont {Das}(2012)}]{DasPRB2012}%
  \BibitemOpen
  \bibfield  {author} {\bibinfo {author} {\bibfnamefont {T.}~\bibnamefont
  {Das}},\ }\href {\doibase 10.1103/PhysRevB.86.054518} {\bibfield  {journal}
  {\bibinfo  {journal} {Phys. Rev. B}\ }\textbf {\bibinfo {volume} {86}},\
  \bibinfo {pages} {054518} (\bibinfo {year} {2012})}\BibitemShut {NoStop}%
\bibitem [{\citenamefont {Izquierdo}\ \emph {et~al.}(2011)\citenamefont
  {Izquierdo}, \citenamefont {Megtert}, \citenamefont {Colson}, \citenamefont
  {Honkim\"{a}ki}, \citenamefont {Forget}, \citenamefont {Raffy},\ and\
  \citenamefont {Com\`{e}s}}]{IzquierdoJPhysChemSolids2011}%
  \BibitemOpen
  \bibfield  {author} {\bibinfo {author} {\bibfnamefont {M.}~\bibnamefont
  {Izquierdo}}, \bibinfo {author} {\bibfnamefont {S.}~\bibnamefont {Megtert}},
  \bibinfo {author} {\bibfnamefont {D.}~\bibnamefont {Colson}}, \bibinfo
  {author} {\bibfnamefont {V.}~\bibnamefont {Honkim\"{a}ki}}, \bibinfo {author}
  {\bibfnamefont {A.}~\bibnamefont {Forget}}, \bibinfo {author} {\bibfnamefont
  {H.}~\bibnamefont {Raffy}}, \ and\ \bibinfo {author} {\bibfnamefont
  {R.}~\bibnamefont {Com\`{e}s}},\ }\href {\doibase 10.1016/j.jpcs.2010.10.055}
  {\bibfield  {journal} {\bibinfo  {journal} {J. Phys. Chem. Solids}\ }\textbf
  {\bibinfo {volume} {72}},\ \bibinfo {pages} {545} (\bibinfo {year}
  {2011})}\BibitemShut {NoStop}%
\bibitem [{\citenamefont {Chabot-Couture}()}]{ChabotCouturePreprint}%
  \BibitemOpen
  \bibfield  {author} {\bibinfo {author} {\bibfnamefont {G.}~\bibnamefont
  {Chabot-Couture}},\ }\href@noop {} {}\bibinfo {note} {Ph.D thesis, Stanford
  University (2010).}\BibitemShut {Stop}%
\end{thebibliography}%

\end{document}